\documentclass[10pt,journal,letterpaper,compsoc, onecolumn]{IEEEtran}
\usepackage{graphicx}
\usepackage{caption}
\usepackage{subcaption}
\usepackage{color}
\usepackage{subfig}
\usepackage{fancyhdr}

\fancypagestyle{empty}{%
  \fancyhf{}
  \fancyhead[L]{\small {\color{blue} This is a preprint of an article accepted for publication in \textit{Social Network Analysis and Mining} copyright @ 2013 (Springer). The final publication is available at link.springer.com.}}
}

\newenvironment{myindentpar}[1]%
 {\begin{list}{}%
         {\setlength{\leftmargin}{#1}}%
         \item[]%
 }
 {\end{list}}

\begin{document}
\pagenumbering{gobble}

\title{Tweet, but Verify: Epistemic Study of Information Verification on Twitter}

\author{
  Arkaitz Zubiaga$^1$, Heng Ji$^2$ \\
  $^1$ Computer Science Department \\
  Queens College and Graduate Center \\
  City University of New York \\
  New York, NY, USA \\
  $^2$ Computer Science Department \\
  Rensselaer Polytechnic Institute \\
  Troy, NY, USA \\
  {\tt arkaitz@zubiaga.org, jih@rpi.edu} \\
}

\IEEEcompsoctitleabstractindextext{%
\begin{abstract}
  While Twitter provides an unprecedented opportunity to learn about breaking news and current events as they happen, it often produces skepticism among users as not all the information is accurate but also hoaxes are sometimes spread. While avoiding the diffusion of hoaxes is a major concern during fast-paced events such as natural disasters, the study of how users trust and verify information from tweets in these contexts has received little attention so far. We survey users on credibility perceptions regarding witness pictures posted on Twitter related to Hurricane Sandy. By examining credibility perceptions on features suggested for information verification in the field of Epistemology, we evaluate their accuracy in determining whether pictures were real or fake compared to professional evaluations performed by experts. Our study unveils insight about tweet presentation, as well as features that users should look at when assessing the veracity of tweets in the context of fast-paced events. Some of our main findings include that while author details not readily available on Twitter feeds should be emphasized in order to facilitate verification of tweets, showing multiple tweets corroborating a fact misleads users to trusting what actually is a hoax. We contrast some of the behavioral patterns found on tweets with literature in Psychology research.
\end{abstract}

}

\maketitle

\IEEEdisplaynotcompsoctitleabstractindextext

\IEEEpeerreviewmaketitle

\thispagestyle{empty}

\section{Introduction}

As social media is gaining more and more importance in our daily lives for information sharing and consumption, Twitter has become the quintessential platform to follow and learn about breaking news and ongoing events \cite{zubiaga2013curating}. However, the fast pace of the stream of tweets not only allows to keep up with current events, but also includes fake reports that often challenge identification of accurate information to get rid of hoaxes. This endangers the trustability of tweets when following events on Twitter, as the veracity of some pieces of information can seem questionable on many occasions.

Assessing the accuracy of information on Twitter becomes more challenging for breaking news and ongoing events. Information about current events often appears on Twitter much earlier than it is reported by news media or documented on websites such as Wikipedia \cite{kwak2010twitter}, making it the only source available by the time and thus reducing chances of contrasting it with other media. Some information is not easily verifiable as quickly as Twitter reacts, e.g., pictures shared by alleged witnesses during a natural disaster can hardly be verified in that moment unless you are in the location of the event and can eyewitness it too. While some information needs time to be verified, users will keep commenting on and sharing tweets, so how and what information is presented to users with each tweet is important to determine how it will be perceived by users and subsequently spread.

While previous research on credibility perceptions of tweets studied what features catch users' attention and condition their decisions when assessing the credibility of a tweet \cite{morris2012tweeting}, the accuracy of those perceptions has not been studied yet. The accuracy of credibility perceptions is what ultimately defines how well the users do when verifying a piece of information \cite{fallis2004verifying}. We are interested in finding out how the impression on different features of a tweet helps or harms toward assessing its veracity. This would allow to present tweets in a more convenient way to avoid hoaxes. For instance, the author has shown to be a feature that users strongly rely on when building credibility perceptions on a tweet; how does this strong reliance lead users to making a good decision on its veracity? To investigate the accuracy of credibility perceptions, we survey users on the credibility of separate features of tweets, hiding the rest of the tweet, and compare their perceptions to final evaluations from professionals. Interestingly, research in Psychology's sub-field of Epistemology has long studied how humans build our beliefs, and what are the factors that make us trust some piece of information to a lesser or greater extent \cite{hume2001enquiry}. We rely on these as a background to define the survey.

Specifically, we analyze the accuracy of credibility perceptions on different features of witness pictures posted on Twitter during Hurricane Sandy's impact and aftermath in the East Coast of the United States in October 2012\footnote{https://en.wikipedia.org/wiki/Hurricane\_Sandy}. Being riddled with fake pictures both from people tweeting jokes and from others trying to gain popularity thanks to hoaxes, it presents an ideal context for the purposes of our study. We analyze credibility ratings provided by Amazon Mechanical Turk\footnote{http://www.mturk.com/} (AMT) workers on different features of the tweets with those pictures (i.e., as previously suggested in epistemology: authority, plausibility, corroboration, and presentation). We evaluate their assessments identifying real and fake pictures from those features, and compare to verification performed by experts, who carefully analyzed pictures using sophisticated techniques. Our study sheds light on the accuracy of credibility perceptions from tweets, finding (i) features that can improve accuracy of credibility perceptions, especially author details currently not readily available on Twitter's feeds, (ii) features that may harm accuracy, such as looking at writing and spelling of a tweet, and (iii) other factors that influence users' perceptions, for instance when repeated exposure of the same hoax received from different authors leads users to mistakenly getting convinced about its veracity. We also discuss the implications of our study for an effective development of a user interface focused on verification of tweets, and posit open research issues that may benefit from further research, such as computation of non-straightforward features useful for verification purposes.

\section{Accuracy of Information and Epistemology}

\subsection{Accuracy of Information}

Accuracy of information is paramount in our society. Inaccurate information can have bad consequences both when the source is well-intended and makes a honest mistake (misinformation) or it is actually intended to deceive (disinformation). It may lead people to making important decisions that can cause serious harm to their health and their finances \cite{fallis2004verifying}. The concern about inaccuracy of information increased with the popularity gain of the Internet, as well as later with social media, which somewhat made it easier to quickly spread wrong information to a large audience \cite{koohang2003misinformation}.

Besides the easiness of spreading inaccurate information to a large number of users on social media sites like Twitter, the fact of disseminating information in real-time poses an additional challenge. In the context of fast-paced events, Twitter has shown to be often the first source to make some breaking news and event announcements \cite{hu2012breaking}, and thus it becomes especially challenging to assess the veracity of a tweet if it cannot be contrasted anywhere else. Moreover, Twitter can also be characterized as a media where users tend to spread hoaxes either for making fun or for trying to gain popularity \cite{hermida2012tweets}. Information shared on Twitter can be easily verifiable in some cases, e.g., when following a soccer game and someone reports a goal scored by one of the teams, we may resort to other sources, or check whether others are reporting the same event. However, there are other kinds of information that are much more difficult to verify, e.g., when someone shares a picture allegedly taken in their neighborhood during a natural disaster. Unless you are around in that moment, or have any other evidence to certify that the picture is real, you will have to trust your own judgment by putting together all the information and background knowledge at your fingertips.

\subsection{Epistemology}
\label{ssec:epistemology}

As a branch of psychology that has been studying humans' understanding and reasoning of acquired knowledge for so many years, we resort to epistemology \cite{piaget1972psychology,goldman1986epistemology}. Besides the nature of knowledge itself, epistemology also studies how knowledge connects to other concepts such as truth or belief. The epistemology of testimony is important because a large amount of the information that we have about the world comes from others rather than from direct observation \cite{lipton1998epistemology}, which forces us to build our beliefs from confidence and credibility perceptions. Moreover, human culture strongly depends on people passing on information, and trusting what others share with them. Resorting to epistemology for the study of information credibility and diffusion on the Web has been considered before by psychologists. For instance, \cite{matheson2004weblogs} performed an epistemic study of news diffusion of blogs as compared to traditional journalism, while \cite{fallis2008toward} relied on epistemology to suggest guidelines to assess information credibility on Wikipedia. Epistemology has not been applied to Twitter before to the best of our knowledge.

Epistemology is a field that conveniently fits with our needs, since it has long studied how we build our beliefs from the knowledge we acquire from our environment. Some fundamental thoughts in this regard include David Hume's classical book ``An Enquiry Concerning Human Understanding'' \cite{hume2001enquiry}, written in 1748. More recently, researchers have resorted to his thoughts to define how we build our beliefs from information sources in the modern era \cite{goldman1999knowledge,fallis2004verifying,fallis2008toward}.

Among those studies, we resort to the guidelines defined by Fallis \cite{fallis2004verifying}. Fallis puts together some of Hume's thoughts and additions by \cite{goldman1999knowledge} to explain how classical theories in human understanding of knowledge can be applied to verify modern information pieces in libraries, newspapers, or the Web. Fallis states that humans build our trust on certain information by putting together a set of characteristics that constitutes and surrounds the information in question. Fallis' arguments can be explained with the following example. Either when checking books in a library, or when searching certain information on the Web, let us think of someone who wants to know how many floors the Empire State Building has. After some search, they find a piece of information stating that ``The Empire State Building, which is located in San Francisco, California, has 102 floors''. From this statement, they are very unlikely to trust that the number of floors is correct. Regardless of the accuracy of the number of floors, the mistake in the location of the building --which is actually located in New York City-- makes it a suspicious source that most probably they will not trust. In the end, the perception of information credibility is built from the aggregations of perceptions of the credibility of different features.

In the process of verifying information, Fallis lists a set of four features that assist humans perform assessments as accurate as possible: (i) authority, (ii) plausibility and support, (iii) independent corroboration, and (iv) presentation. We rely on these four features to study how they can be applied to information verification on Twitter, and discuss how our findings can help develop and design effective systems with the end of leading users to accurate information.

In this paper, we describe how these four features can be applied to information verification on Twitter. We survey users to capture their credibility perceptions on witness pictures shared in the context of Hurricane Sandy in late 2012, as a fast-paced event where pictures could hardly be contrasted immediately, and measure their accuracy by contrasting their perceptions to final assessments performed by experts.

\section{Epistemic Study of Tweets}

In this section, we describe the data gathering process, the survey we conducted for collecting credibility perceptions on tweets, and analyze these perceptions and their accuracy.

\subsection{Data Collection}
\label{ssec:data-collection}

We collected tweets from Twitter's streaming API by tracking the terms 'sandy' and 'hurricane' (which implicitly retrieves also tweets containing the hashtags '\#sandy' and/or '\#hurricane'). Note that by following these two terms, we broadened the scope of users who tweeted about the event beyond those that only used the hashtags. We chose to track the terms 'hurricane' and 'sandy' as they were the predominant terms being spread both on Twitter and on other channels such as TV and newspapers early on after the hurricane formed and began to approach. We tracked tweets from October 29 to November 1, 2012, that is, while the hurricane was hitting the East Coast of the United States, as well as in the aftermath. That included as many as 14.9 million tweets. For the purposes of this research study, we created groups of tweets, where tweets in each group were pointing to the same picture --i.e., pictures whose md5 checksums were identical--, and removed tweets that did not include a picture. Still, after the md5 checksum verification, some of the remaining pictures were slightly edited versions of others, which we left ungrouped. All the pictures considered for this study were uploaded either to Twitter's own photo service\footnote{http://pic.twitter.com/} or to Instagram\footnote{http://www.instagram.com/}. To facilitate carrying out the survey with US-based users, we considered the groups of tweets in which the source tweet (i.e., the first tweet) was written in English\footnote{We filter by language of the source tweet, because that is the tweet we show during the surveys.}. We considered pictures with at least 100 tweets associated, so they were popular and thus likely to be later analyzed by a number of professionals and discussed on the Web. From this process, we got a final set of 332 pictures posted on Twitter associated with Hurricane Sandy. Note that those 100+ tweets include retweets\footnote{Retweets were identified as tweets containing the field ``retweeted\_status'' pointing to the original tweet.}, and new original tweets that borrowed the picture in some cases.

To label each of these 332 pictures as real or fake, we relied on decisions performed later by picture verification professionals. Hurricane Sandy became extremely popular while it was hitting the East Coast of the United States, so it attracted the attention of many journalists, photographers, and publishers who carefully investigated each and every picture posted on social media. In many cases, verification of pictures was crowdsourced and performed through collaboration among professionals, so it is relatively easy to find the final decisions on the Web. We used two search methods to look for professional assessments. On one hand, we looked at several lists posted on the Web\footnote{e.g., http://www.theatlantic.com/technology/archive/2012/10/sorting-the-real-sandy-photos-from-the-fakes/264243/}, complemented with an image search by using Google's Image Search\footnote{http://images.google.com/} to search for professional evaluations on a specific picture. Professionals carefully analyzed popular pictures posted on Twitter by looking at not only whether the pictures made sense in the context, location and time of the hurricane, or searching for previous uses of the pictures on the Web, but also using other sophisticated techniques such as analysis of angles, shadows, and reflections that might have been manipulated \cite{o2012exposing}. After the labeling process, we obtained a dataset comprised of 216 real pictures (65.1\%) and 116 fake pictures (34.9\%).

\begin{figure}
 \centering
 \begin{subfigure}[b]{0.3\textwidth}
  \includegraphics[width=\textwidth]{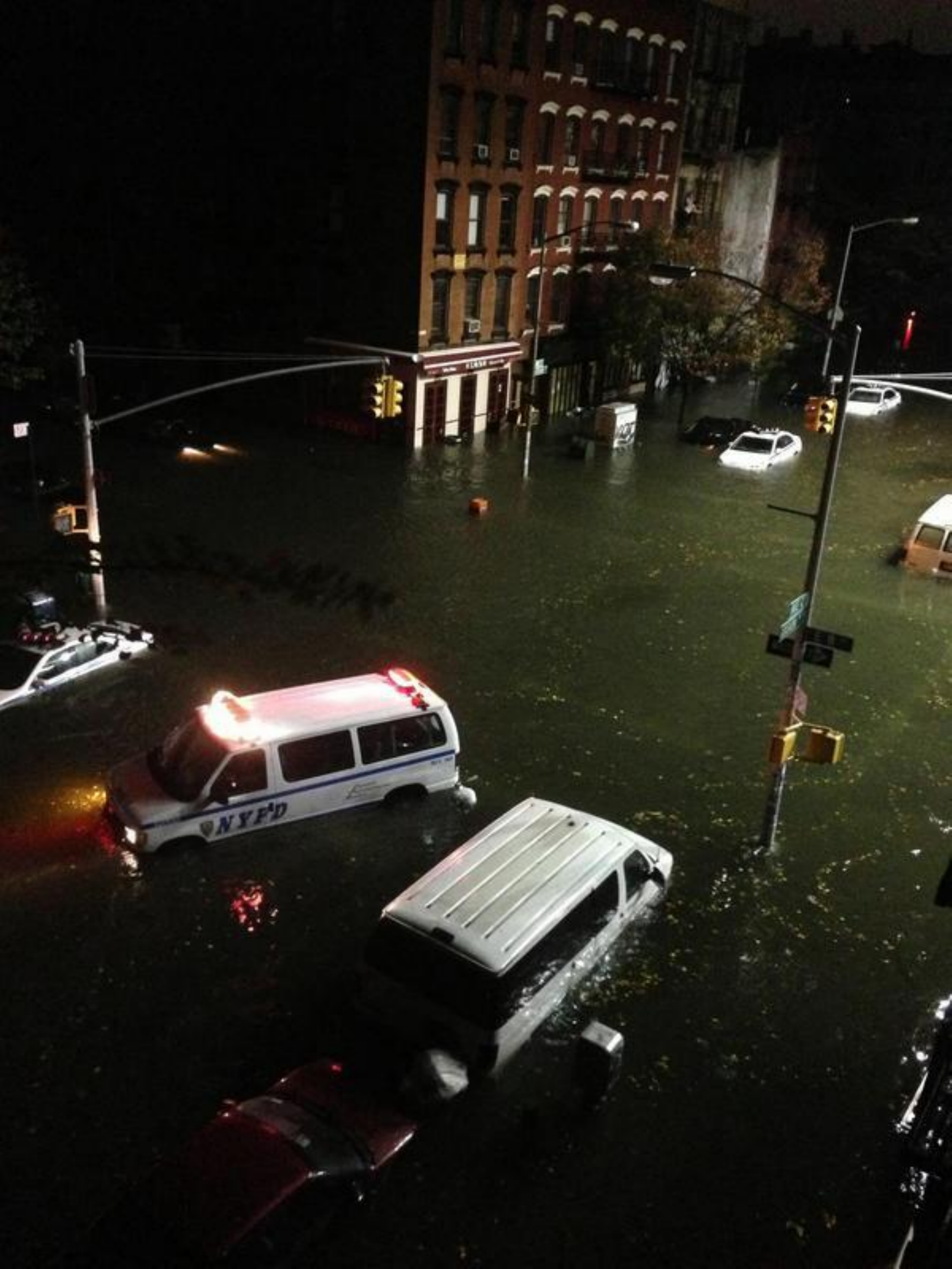}
  \caption{Real}
  \label{fig:real1}
 \end{subfigure}
 \begin{subfigure}[b]{0.3\textwidth}
  \includegraphics[width=\textwidth]{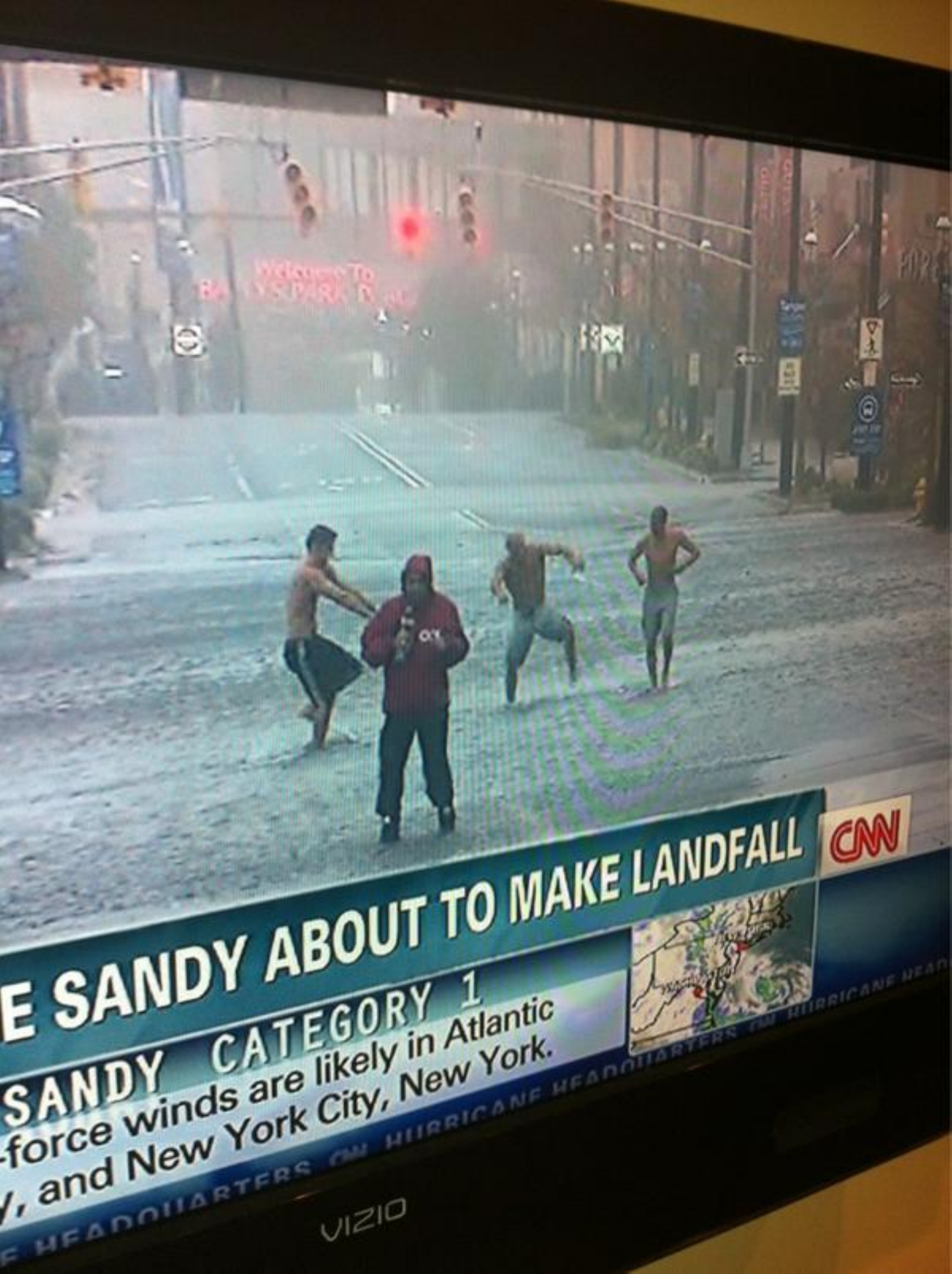}
  \caption{Real}
  \label{fig:real2}
 \end{subfigure}
 \begin{subfigure}[b]{0.3\textwidth}
  \includegraphics[width=\textwidth]{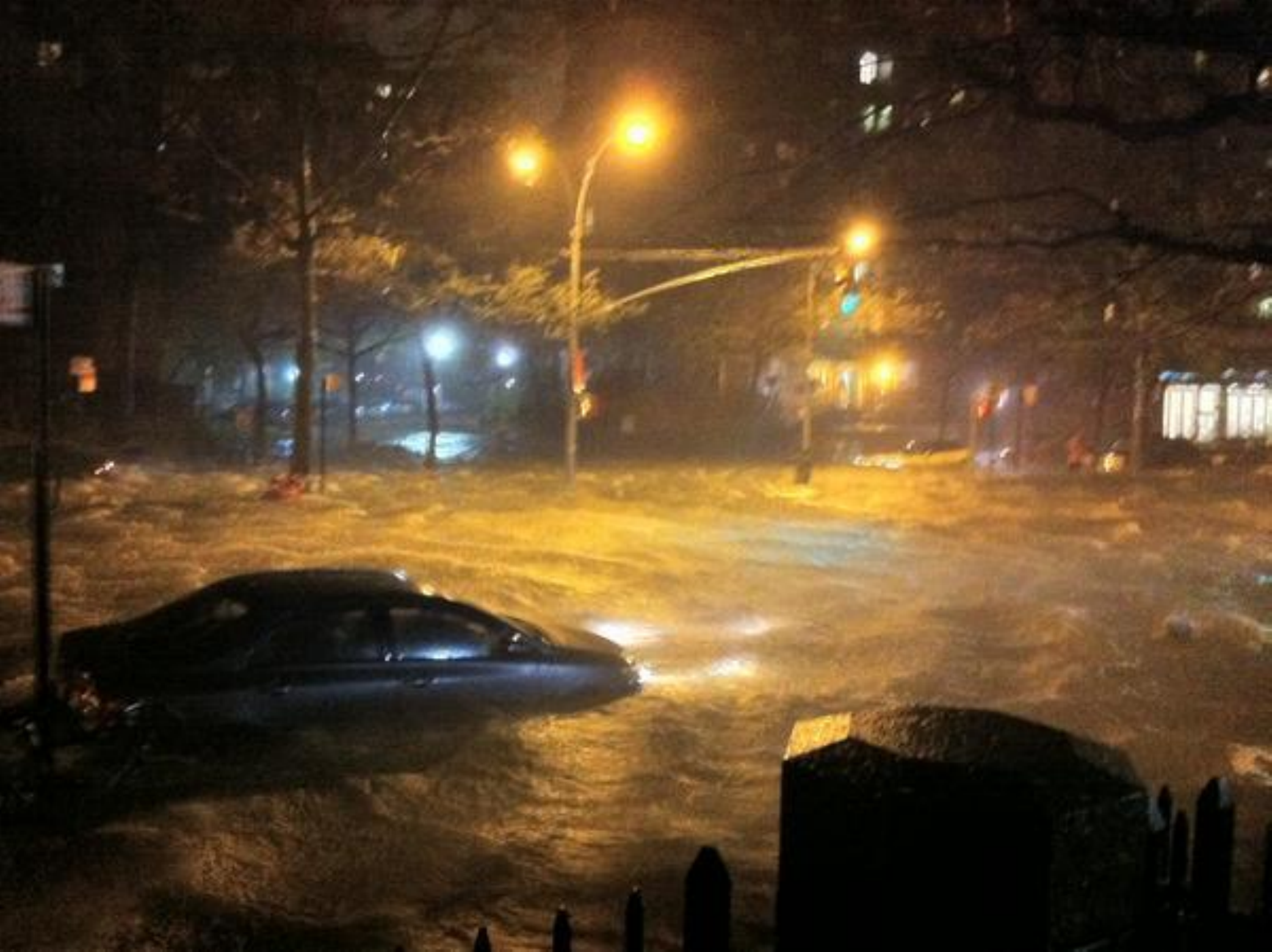}
  \caption{Real}
  \label{fig:real3}
 \end{subfigure}
 \begin{subfigure}[b]{0.3\textwidth}
  \includegraphics[width=\textwidth]{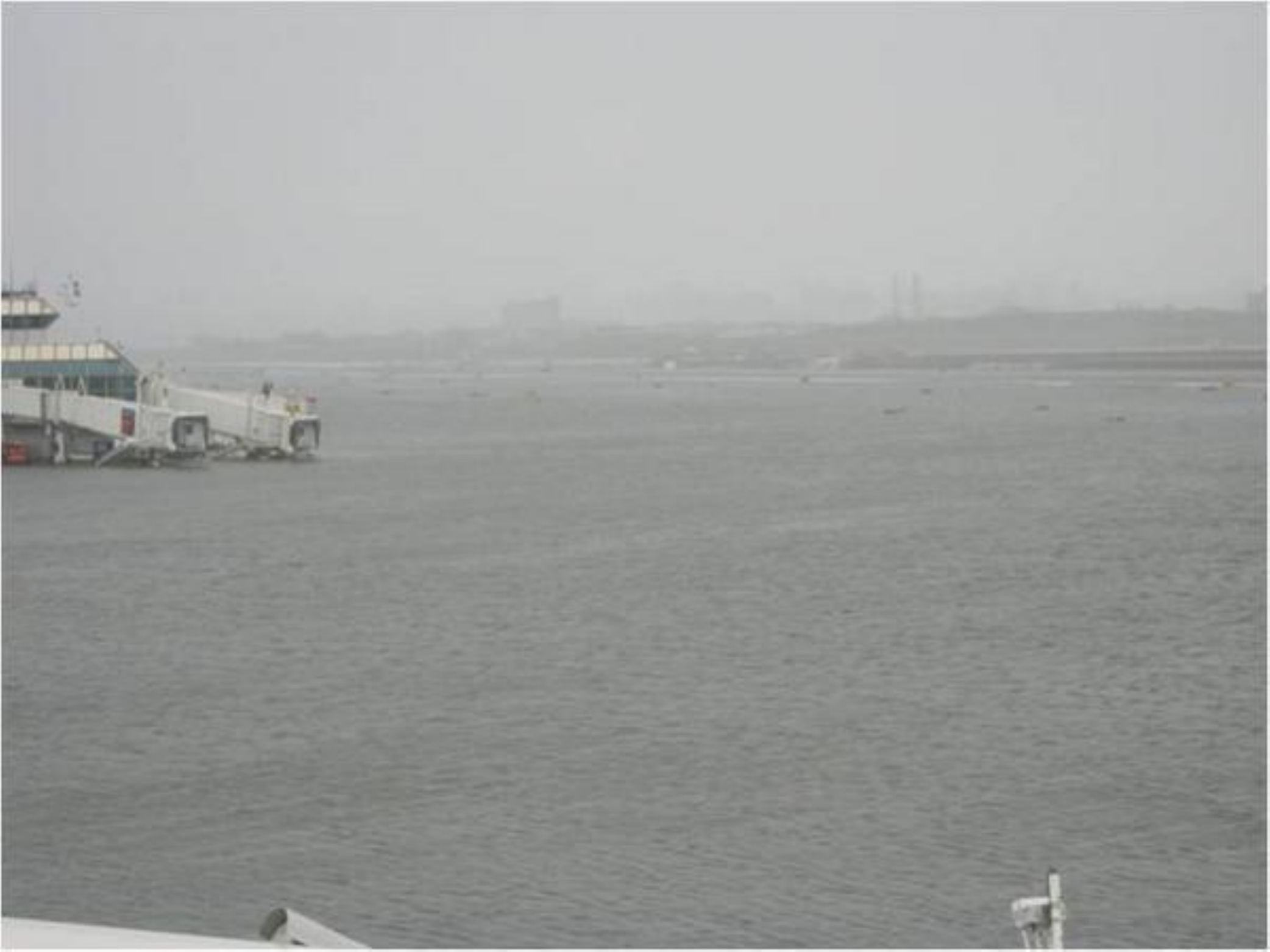}
  \caption{Real}
  \label{fig:real4}
 \end{subfigure}
 \begin{subfigure}[b]{0.3\textwidth}
  \includegraphics[width=\textwidth]{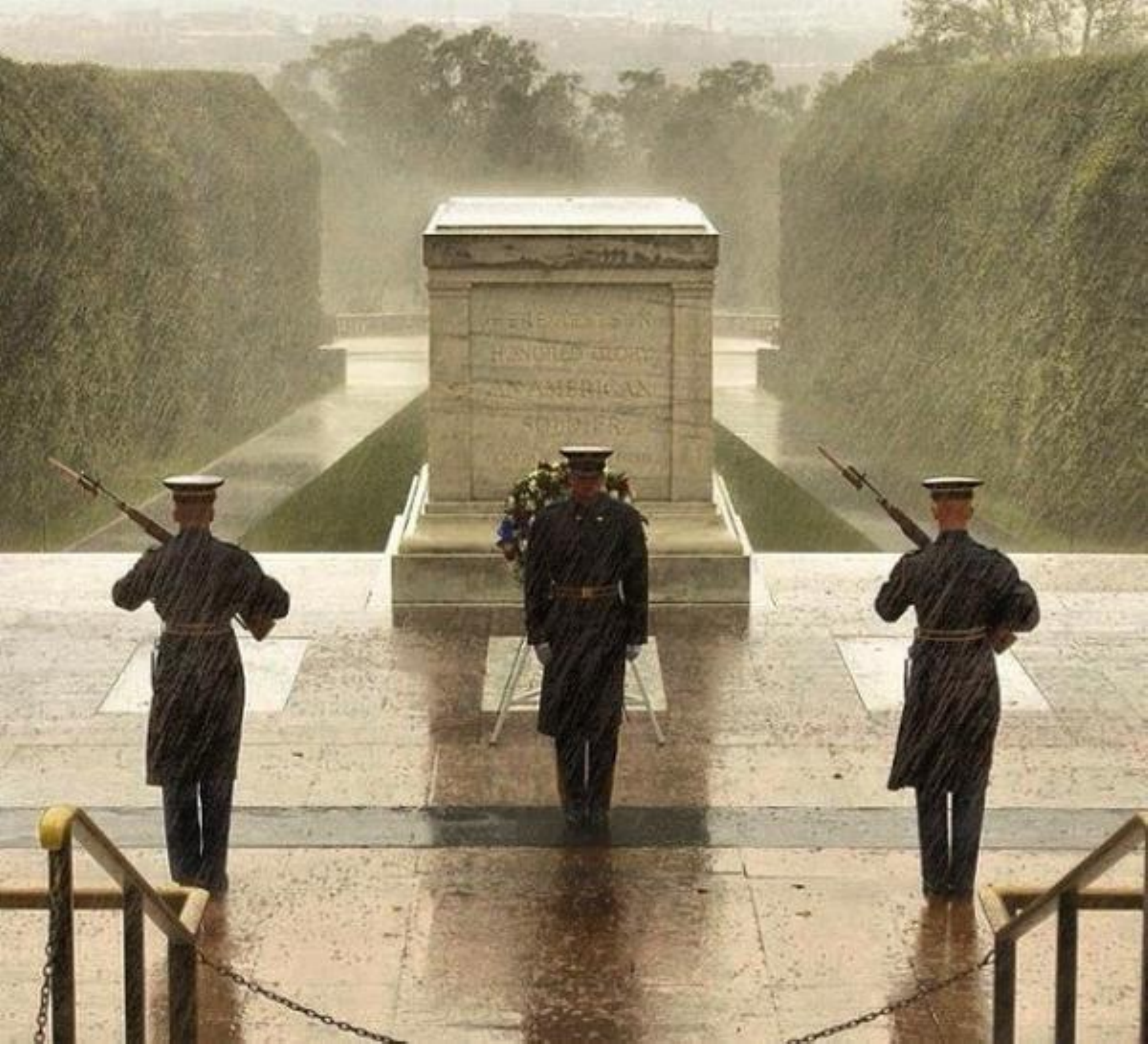}
  \caption{Fake}
  \label{fig:fake1}
 \end{subfigure}
 \begin{subfigure}[b]{0.3\textwidth}
  \includegraphics[width=\textwidth]{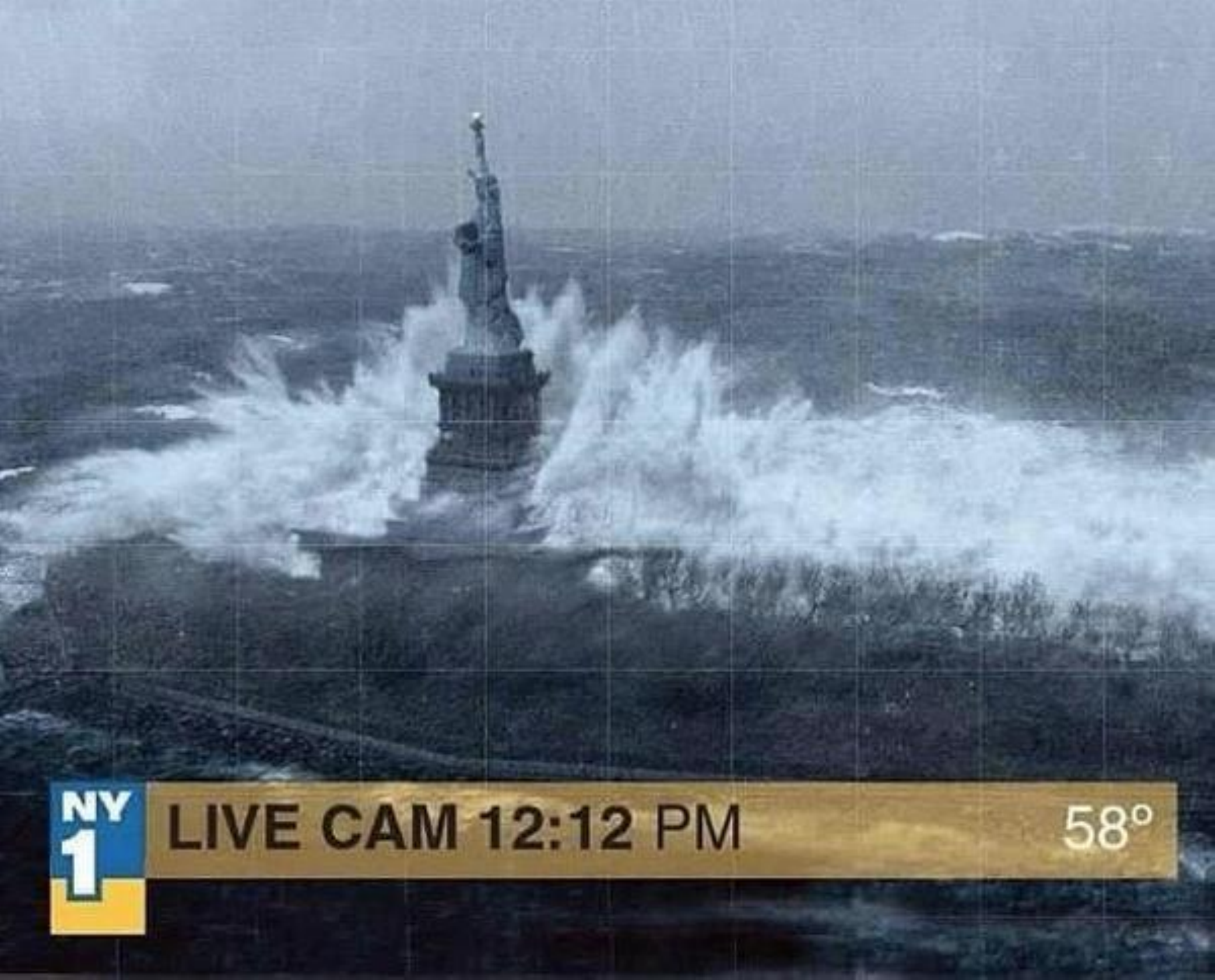}
  \caption{Fake}
  \label{fig:fake2}
 \end{subfigure}
 \begin{subfigure}[b]{0.3\textwidth}
  \includegraphics[width=\textwidth]{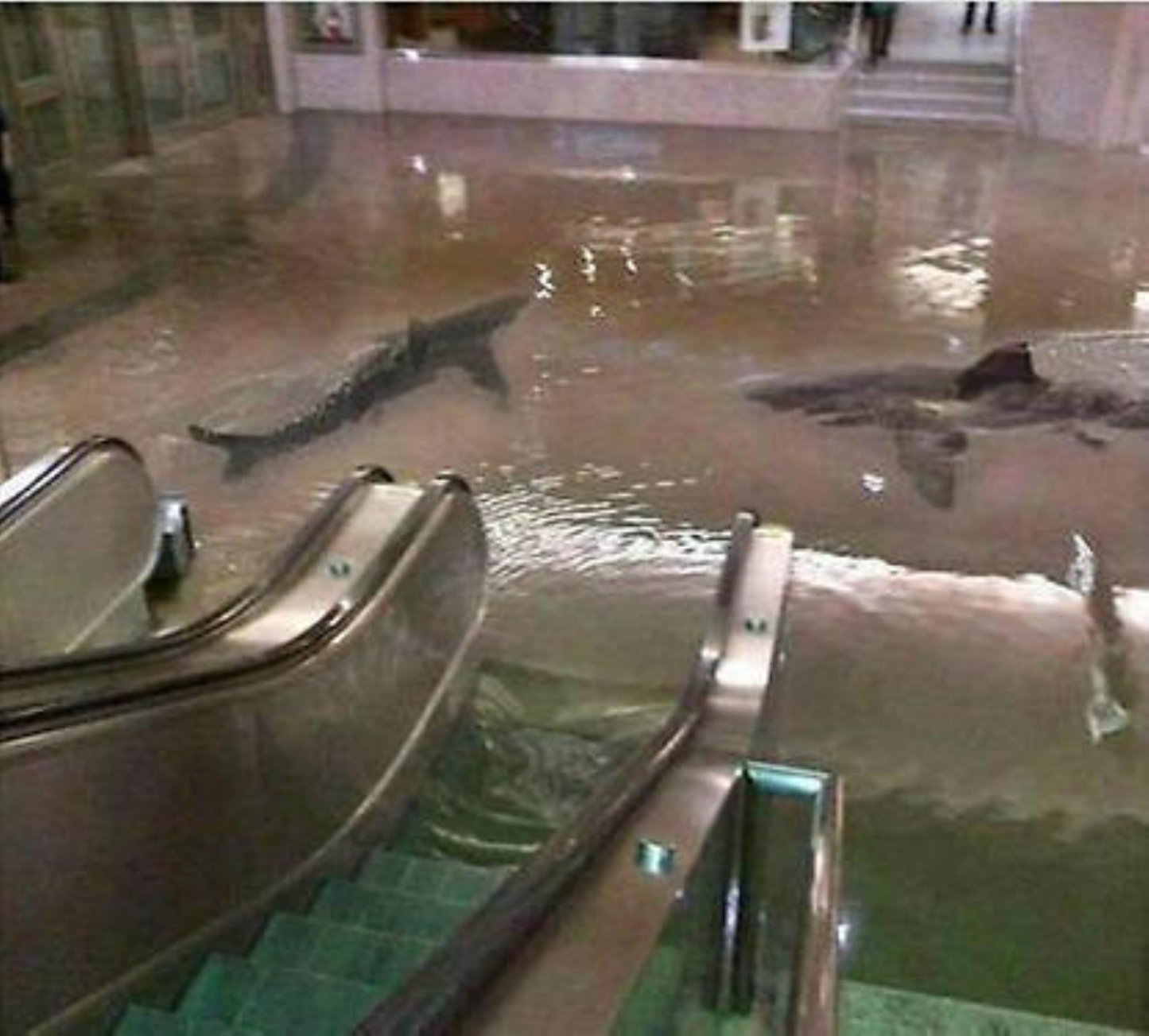}
  \caption{Fake}
  \label{fig:fake3}
 \end{subfigure}
 \begin{subfigure}[b]{0.3\textwidth}
  \includegraphics[width=\textwidth]{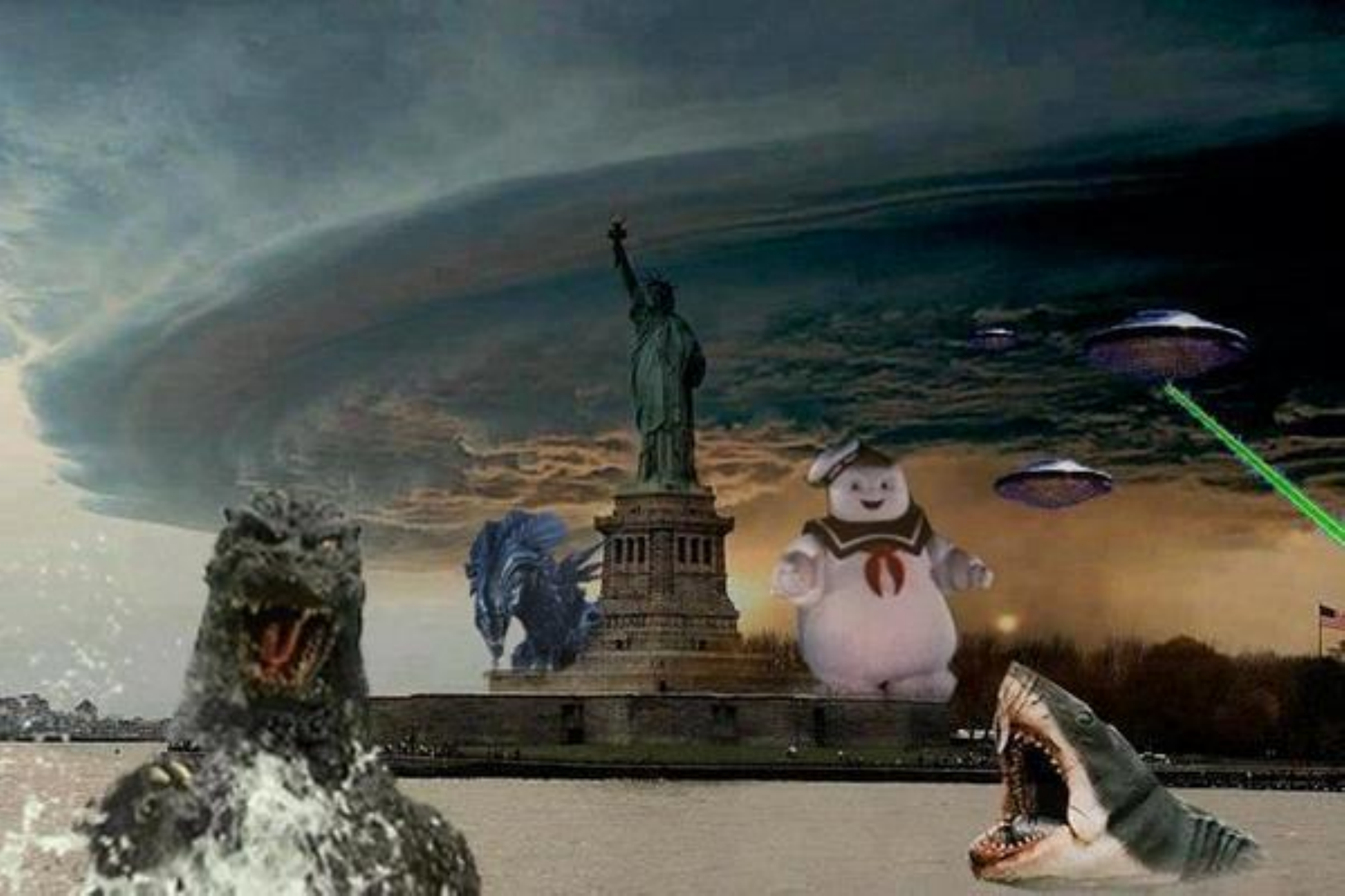}
  \caption{Fake}
  \label{fig:fake4}
 \end{subfigure}
 \caption{Examples of pictures posted on Twitter during the effects of Hurricane Sandy.}
 \label{fig:examples}
\end{figure}

Figure \ref{fig:examples} shows examples of pictures posted during the effects of Hurricane Sandy, and are part of the dataset used in our study. On one hand, pictures \ref{fig:real1}, \ref{fig:real2}, \ref{fig:real3}, and \ref{fig:real4} were actually real, despite they seem certainly questionable. Figures \ref{fig:real1} and \ref{fig:real3} show pictures of flooded streets. One might really doubt if the streets of New York City were in that condition at the moment, whether or not they were really taken in New York City, as well as if the photographer could be in a position to take such pictures. Trusting those requires careful analysis. Figure \ref{fig:real2} shows a picture taken from a TV reporting the news, where the presence of some dancing guys in the back of the reporter question whether the author might have photoshopped it as an attempt at humor. Figure \ref{fig:real4} shows an unprecedented picture of a flooded Laguardia airport where the water's level almost reaches the jet bridges. The difficulty of trusting these pictures increases in an emergency event where the number of witnesses is scarce. On the other hand, pictures \ref{fig:fake1}, \ref{fig:fake2}, \ref{fig:fake3}, and \ref{fig:fake4} turned out to be fake, some of which were initially deemed real, but later exploration clarified their fakeness. Figure \ref{fig:fake4} shows one of the few cases in the dataset in which the author clearly attempted at humor, photoshopping the picture with noticeably fake images of film characters. However, some of the real pictures did also include hints of humor that could question their veracity, as it happens with Figure \ref{fig:real2}. The presence of humor in both real and fake pictures makes verification more challenging. Most of the pictures, though, intended to deceive with credible situations. Figure \ref{fig:fake1} shows a picture of three soldiers allegedly guarding the Tomb of the Unknowns (located in Arlington, Virginia) while the hurricane was hitting the area. The picture turned out to be taken a month earlier, and so it had nothing to do with the hurricane. Figure \ref{fig:fake2} shows a wave-battered Statue of Liberty in New York as if it were broadcast live by New York's local news broadcasting channel \textit{NY1}. While many might trust this picture in such situation, the picture was actually manipulated after taking a screenshot from the disaster movie \textit{The Day After Tomorrow}. Figure \ref{fig:fake3} shows a picture with two sharks swimming next to escalators, which was a hoax also spread earlier stating that sharks made it into a popular mall in Kuwait. Especially provided that real pictures were also stunning, any of the pictures was questioned at the moment as potentially being faked, and the fact that many attempted to deceive the social media audience made the task of assessing the veracity of pictures more challenging.

As an initial analysis, we looked at the number of tweets and retweets associated with each picture. With this, we wanted to look at whether there was a correlation between the number of tweets and pictures being real or fake, which would suggest that Twitter users did well on spreading just real pictures. However, we noticed that the popularity does not correlate with the truthfulness of the picture in question, and fake pictures had become as popular as real pictures. As shown in Table \ref{tab:fake-real-calculations}, fake pictures present a higher median value, as well as a higher average with a bigger standard deviation. It is hard to know, however, if all the retweets and new tweets of fake pictures occur because users considered they were real, or because they were making fun of or questioning their fakeness. Although the believability of information has sometimes been identified as a factor determining whether it is propagated \cite{cotter2008influence}, the behavior we found on Twitter coincides with \cite{lewandowsky2012misinformation} in that people seem to pass on information that will evoke an emotional response in the recipient, irrespective of the information's truth value. This emotional response can be either guffaw with funny fake pictures, or awe with seemingly real pictures, for example, which might lead users to retweeting in both cases.

\begin{table}[htb]
\begin{center}
 \begin{tabular}{|l|c|c|c|}
  \hline
  & \textbf{Median} & \textbf{Average} & \textbf{Std. Deviation} \\
  \hline
  \textbf{Real} & 216 & 425.7 & 521 \\
  \hline
  \textbf{Fake} & 261 & 670.6 & 1020.8 \\
  \hline
 \end{tabular}
\end{center}
\caption{Median, average and standard deviations in terms of number of shares for real and fake pictures.}
\label{tab:fake-real-calculations}
\end{table}

\subsection{Epistemology of Tweets}

Having collected the data, and defined a ground truth from professional assessments as a reference, we proceeded to set out the user studies to collect credibility perceptions on tweets. Next, we describe the four features suggested by Fallis, which we introduced in Section \ref{ssec:epistemology}, explain how we apply them in the context of Twitter, and detail the surveys conducted on AMT. For each of the Twitter pictures, we rely on the first tweet posting a picture as the source tweet to represent the features. Analogous to Fallis' definition of information source, the first tweet is the one that serves as a source since it first publishes the information and later produces reactions in the form of retweets and new tweets from others. In the event that any of the pictures might be familiar to AMT workers rating their credibility perceptions, we asked them not to rate a tweet if they already knew whether it was real or fake. In separate tasks for each feature, we asked AMT workers to rate each of the features, without having access to the rest of the tweet; later, we ran a complementary task where we asked them to rate tweets as they would see them on Twitter. The workers were given detailed guidelines along with the tweets (or parts of tweets) to be rated, with slight variations depending on the task and the part of the tweet being shown. We showed a set of five tweets (or parts of them) in each task, along with the following guidelines:

\begin{myindentpar}{1cm}
 \textit{You're shown a set of tweets sent during or just after Hurricane Sandy hit the northeast of the US. These tweets include a picture attached to them. The pictures might be real or fake (i.e. manipulated, not from the affected area, etc.). By looking at the whole tweet, we need you to rate how plausible you would consider it in the context of Hurricane Sandy. In the event that any of the tweets/pictures might be familiar to you, don't rate it.}

 \textit{Please, rate from 1 (least plausible) to 5 (most plausible), according to the following:}

 \begin{itemize}
  \item \textit{1: It doesn't look plausible to me, I wouldn't trust.}
  \item \textit{2: It doesn't look plausible, but I wouldn't be 100\% sure.}
  \item \textit{3: It seems suspicious to me, not sure if I would trust or not.}
  \item \textit{4: It looks plausible, but I'm not 100\% sure.}
  \item \textit{5: It looks plausible to me, I would trust.}
 \end{itemize}
\end{myindentpar}

All of the features were rated by 9 workers each, on a likert scale from 1 to 5, defined as follows: (1) not plausible at all, (2) not plausible, but unsure, (3) uncertain, (4) plausible, but unsure, and (5) plausible at all. We define the final ratings for each pair of feature and picture as the average of ratings by the 9 workers.  To measure the inter-rater agreement we use Krippendorff's alpha \cite{krippendorff2012content} over other metrics because of its capacity to consider that some categories are closer to one another (e.g., an inter-rater disagreement between 4 and 5 is much smaller than a disagreement between 1 and 5). We report the strength of agreement from the benchmarks suggested by Landis and Koch \cite{landis1977measurement} and widely adopted afterward \cite{stemler2001overview} for interpreting values of kappa.

\textbf{Feature 1: Authority.} The author of a piece of information is one of the features that determines whether it is likely to be true or false. However, there are lots of factors to look at when analyzing who the author is. There are some that are hardly applicable to Twitter, e.g., the question ``has this author usually provided accurate information in the past?'' is not always easy to answer when following an event on Twitter, since it is very unlikely that the author is known to those who do not follow them. Interestingly, there are other questions that users can try to answer to assess the likelihood of an author to be posting the truth. A remarkable question is: ``does anything suggest that this information source would not provide accurate information in this particular case?''. By looking at some features like the location of an author, the description, or the number of followers, which are not readily available on Twitter's feeds, users can have an impression of how likely the author is to be saying the truth. Another question suggested by Fallis is ``people can determine whether an information source on the Internet is endorsed by others''; although in this case he refers to the number of links pointing to a website, this can be analogous to the number of followers on Twitter. The last question says ``is there any indication that the witness was not in a position to know the fact that she is testifying to?'', which in the context of Twitter can be inferred, when available, from the location and description of the user.

To collect credibility perceptions of users on Twitter authors in the context of Hurricane Sandy, we asked AMT workers to rate the authors who first posted each of the 332 pictures in our dataset. For each of these authors, they were given the user name and full name, profile picture, description, and number of followers and followees. Having that information, and instructed by the above questions, we asked workers to rate according to their perception of how likely an author would have posted a real picture in the context of Hurricane Sandy. The inter-coder agreement when rating authors of tweets was Krippendorff's $\alpha = 0.202$, representing fair agreement.


\textbf{Feature 2: Plausibility and Support.} The second feature that helps separate real from fake information is by looking at how plausible a piece of information is, as well as the reasons offered in support of the information. The degree of credibility perceived from a piece of information is conditioned by the background knowledge of the user reading it. This determines how a user processes the information being read, matching with what they have learned before to be likely to happen in a certain situation. Some information would be credible to the user when either the claim itself or the supporting information given makes sense to the prior knowledge the user has on the topic.

We ran two different tasks on AMT regarding plausibility and support, separating the text and the picture to analyze perceptions in each component of the tweet:

\textbf{Feature 2.1: Text plausibility.} On one hand, we provided AMT workers with just the text of the tweet, with no picture and no author information, and asked them to rate how credible they would consider that the tweet had a real picture associated by following the above guidelines in the context of Hurricane Sandy. The inter-coder agreement when rating plausibility of tweets was Krippendorff's $\alpha = 0.284$, representing fair agreement.


\textbf{Feature 2.2: Picture plausibility.} On the other hand, AMT workers were instructed to rate, following the same guidelines, the picture associated with the tweet. In this case, they did not have access to the text, and rated it by just looking at the picture. The inter-coder agreement when rating pictures was Krippendorff's $\alpha = 0.427$, representing moderate agreement.


\textbf{Feature 3: Independent Corroboration.} The third feature suggested by Fallis to verify the accuracy of information is the complementary information from other sources supporting the claim, i.e., whether or not and how others corroborate the information. The idea behind this is that it is much more likely that one individual will intend to deceive others, but it is less likely to have several sources who attempt to deceive, and all of them do it in exactly the same way. However, getting the same information from different sources is not always an indication that their information is accurate, and it is important to check whether those sources are independent from each other. Unfortunately, determining whether corroboration claims are independent from one another is up to the reader, and might not always be easy on Twitter, unless a user is mentioned as a source.

In this case, complementing the information from source tweets showed for previous tasks, we extracted all independent tweets associated with each picture. By independent tweets, we refer to those that are not retweets of others, and are written as new tweets. Since this feature depends on the availability of corroborating tweets, we could only get ratings for 161 tweets (48.5\% of the whole set); the other tweets did not have any corroborating tweets, but just retweets of the same tweet, which provided no new information. For these 161 tweets, AMT workers were asked to rate, after seeing first only the original source tweet, how credible it was for them that the set of corroborating tweets had a real picture associated. This task is similar to the feature 2.1, i.e., plausibility of the source tweet, with the addition that workers could see more tweets providing complementary information. We compare these two related features to understand the effect of corroborating tweets. The inter-coder agreement when rating corroboration was Krippendorff's $\alpha = 0.177$, representing slight agreement.


\textbf{Feature 4: Presentation.} The fourth and last feature is complementary to the information regarding authorship and content considered by the above features. Presentation does not only refer to the design used to show the information, which does not make any difference among tweets, as the design is predefined. Interestingly, and applicable to Twitter, Fallis emphasizes the importance of how the author testifies as an indicator of its reliability. This applies to the use of appropriate writing, spelling and grammar, which are the three factors that we considered for presentation with regard to tweets.

We asked AMT workers to rate each of the 332 tweets according to their level of presentation, showing only the text of the tweet. Following the guidelines above, we asked them to rate high a tweet when it was well written with formal writing and no swear words, and used correct spelling and grammar. The inter-coder agreement when rating presentation was Krippendorff's $\alpha = 0.360$, representing fair agreement.


\textbf{Whole tweet.} To complete the study, we aimed to compare the accuracy of credibility perceptions on the four separate features described above to the accuracy of perceptions of the whole tweet as shown on Twitter. We also asked AMT workers to rate tweets by looking at the details they would see on Twitter or most major applications. For this task, AMT workers were shown tweets containing the profile picture and username of the author, text, and the picture posted, as they would usually see them. They rated credibility perceptions on each of the 332 tweets, within the same range from 1 to 5, as for the features above. The inter-coder agreement when rating whole tweets was $\alpha = 0.391$, representing fair agreement.

\subsection{Results}

Now, we analyze the final ratings, computed as the arithmetic mean of the 9 ratings in each case. For each picture, we obtained 6 ratings: authority, text plausibility, picture plausibility, corroboration\footnote{with the exception that corroboration is not always available}, presentation, and the whole tweet. Since we have the final assessments given by professionals for each of the pictures, we are interested in analyzing not only how features relate between them, but also in studying how accurate workers were when rating each of the features, i.e., verifying tweets.

\begin{figure*}[t]
 \centering
 \begin{subfigure}[b]{0.3\textwidth}
  \centering
  \includegraphics[width=\textwidth]{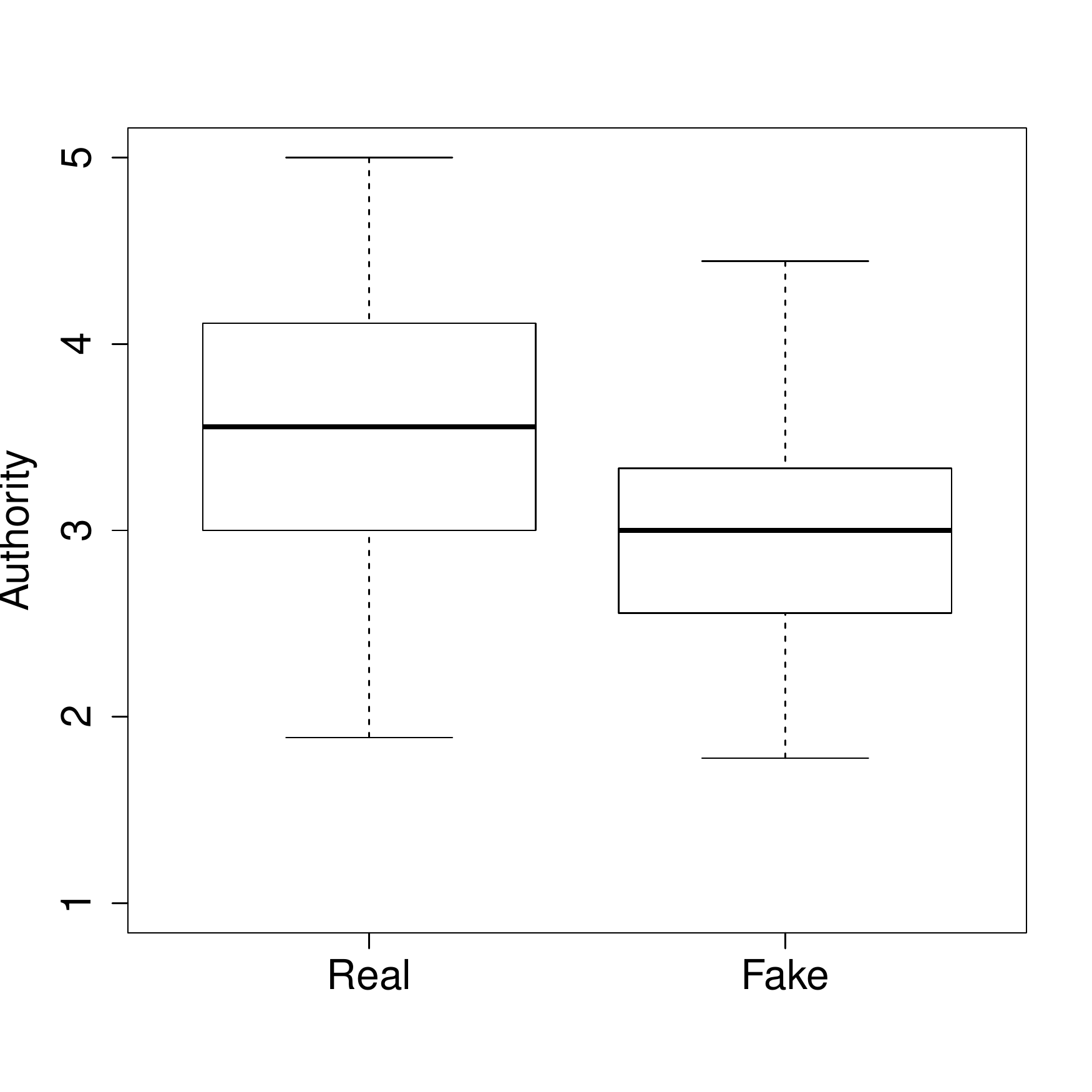}
  \caption{Authority}
  \label{fig:a-box}
 \end{subfigure}
 \begin{subfigure}[b]{0.3\textwidth}
  \centering
  \includegraphics[width=\textwidth]{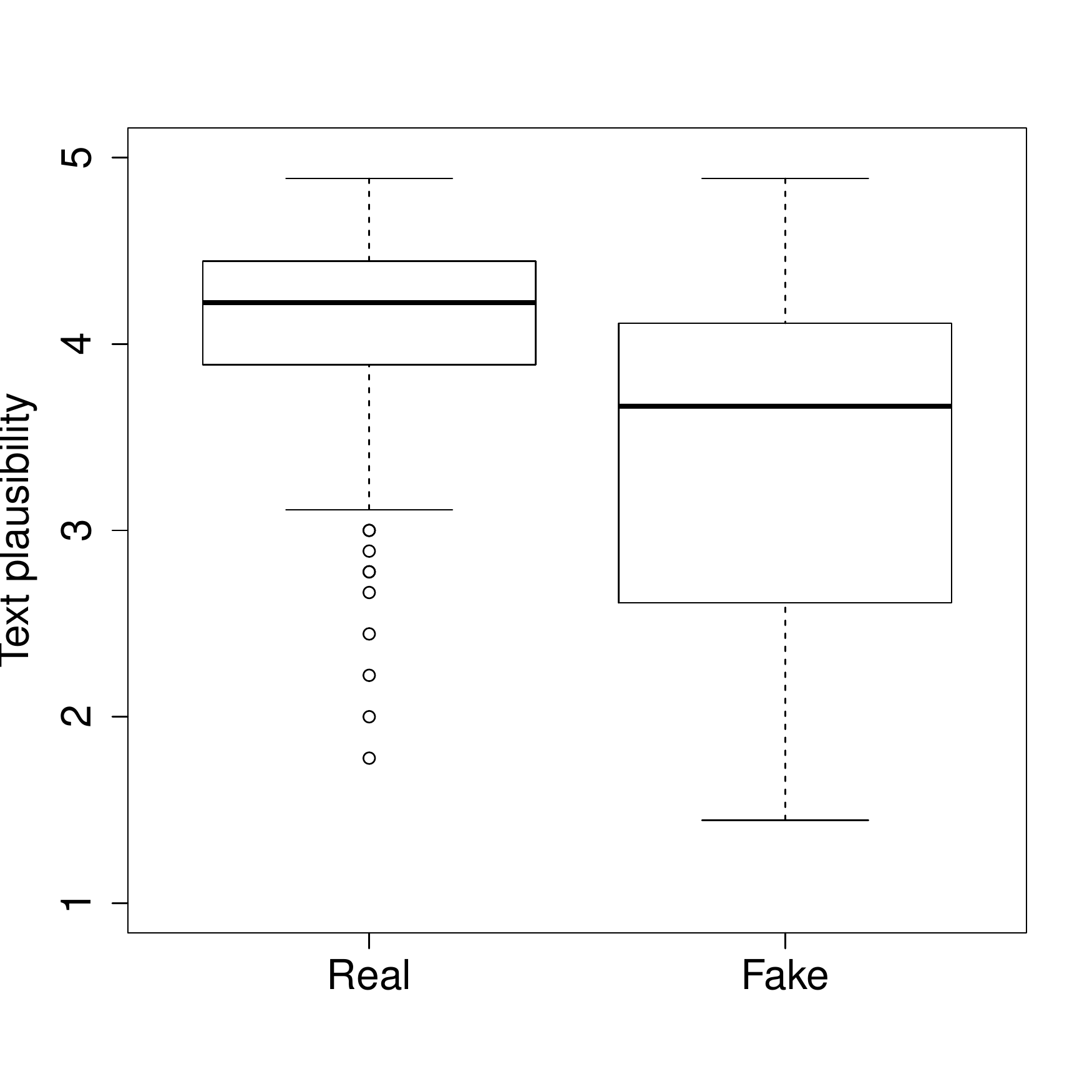}
  \caption{Text plausibility}
  \label{fig:d-box}
 \end{subfigure}
 \begin{subfigure}[b]{0.3\textwidth}
  \centering
  \includegraphics[width=\textwidth]{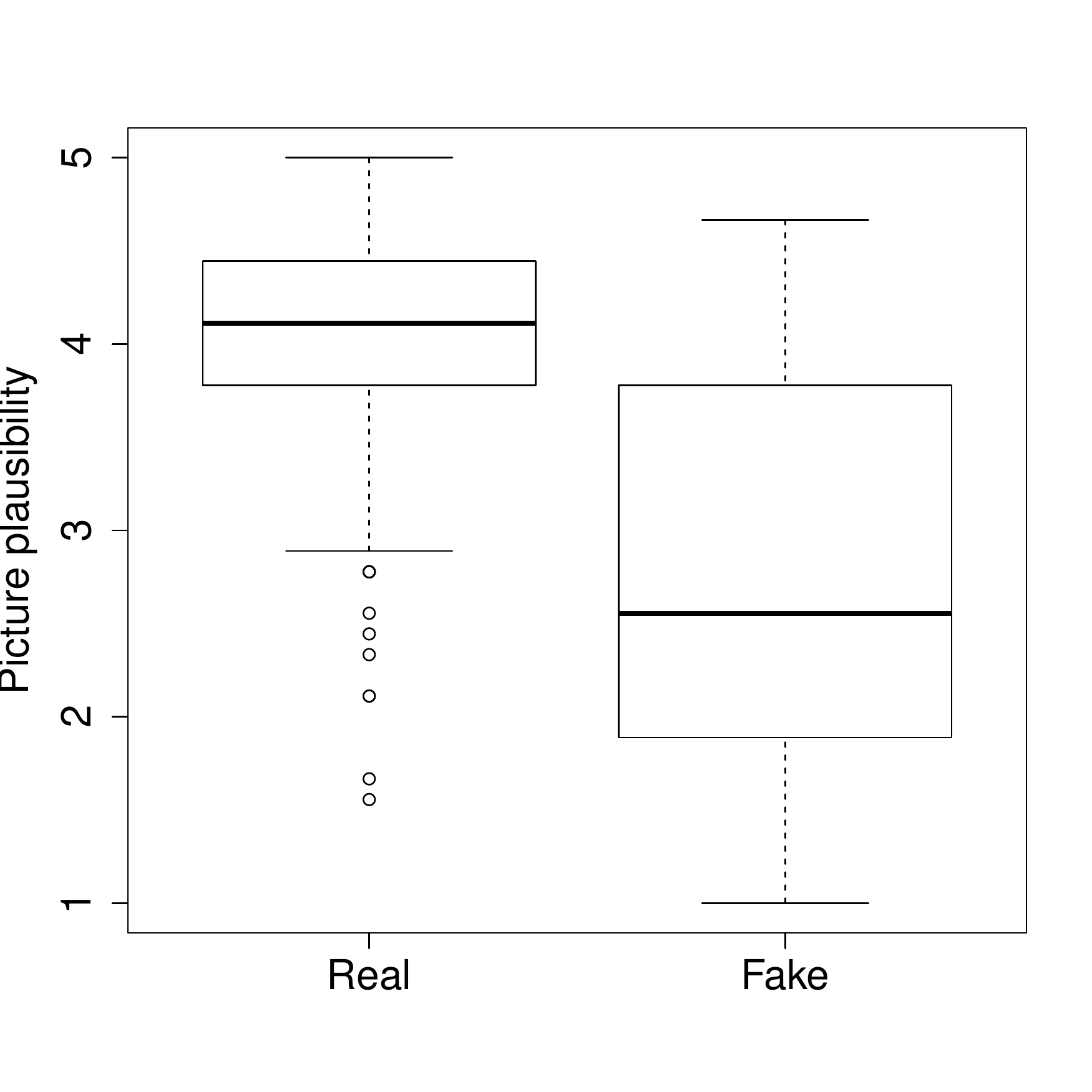}
  \caption{Picture plausibility}
  \label{fig:c-box}
 \end{subfigure}
 \begin{subfigure}[b]{0.3\textwidth}
  \centering
  \includegraphics[width=\textwidth]{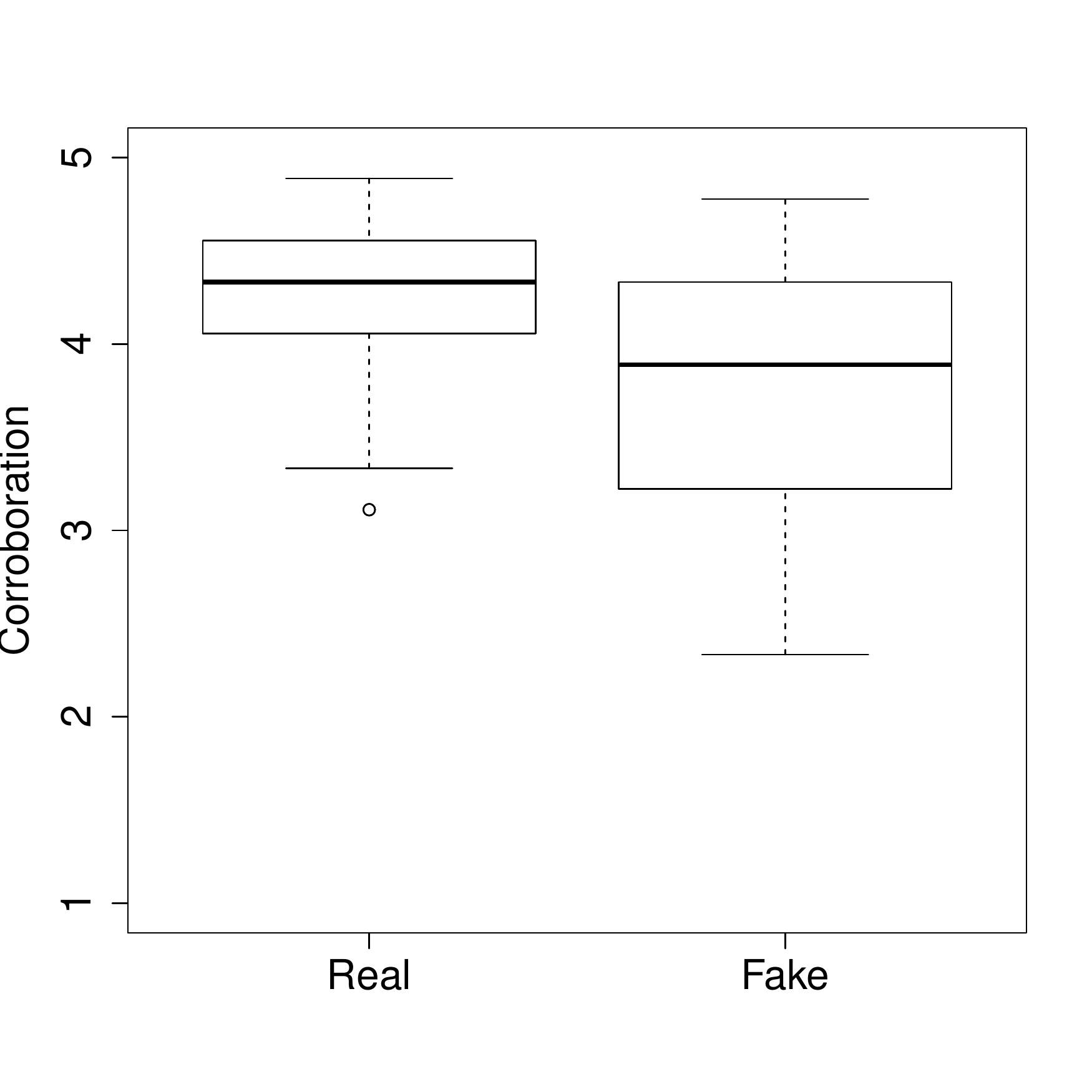}
  \caption{Corroboration}
  \label{fig:b-box}
 \end{subfigure}
 \begin{subfigure}[b]{0.3\textwidth}
  \centering
  \includegraphics[width=\textwidth]{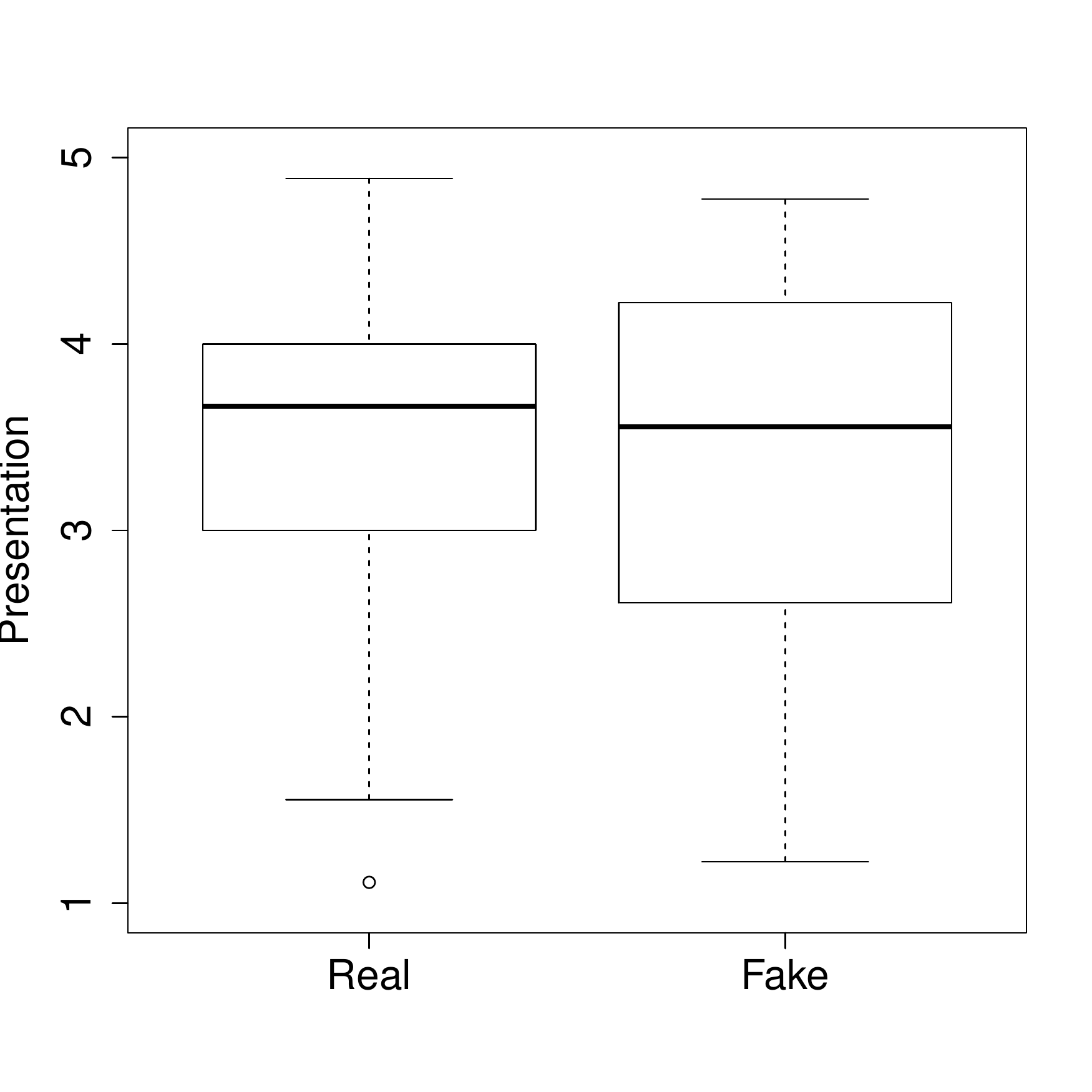}
  \caption{Presentation}
  \label{fig:e-box}
 \end{subfigure}
 \begin{subfigure}[b]{0.3\textwidth}
  \centering
  \includegraphics[width=\textwidth]{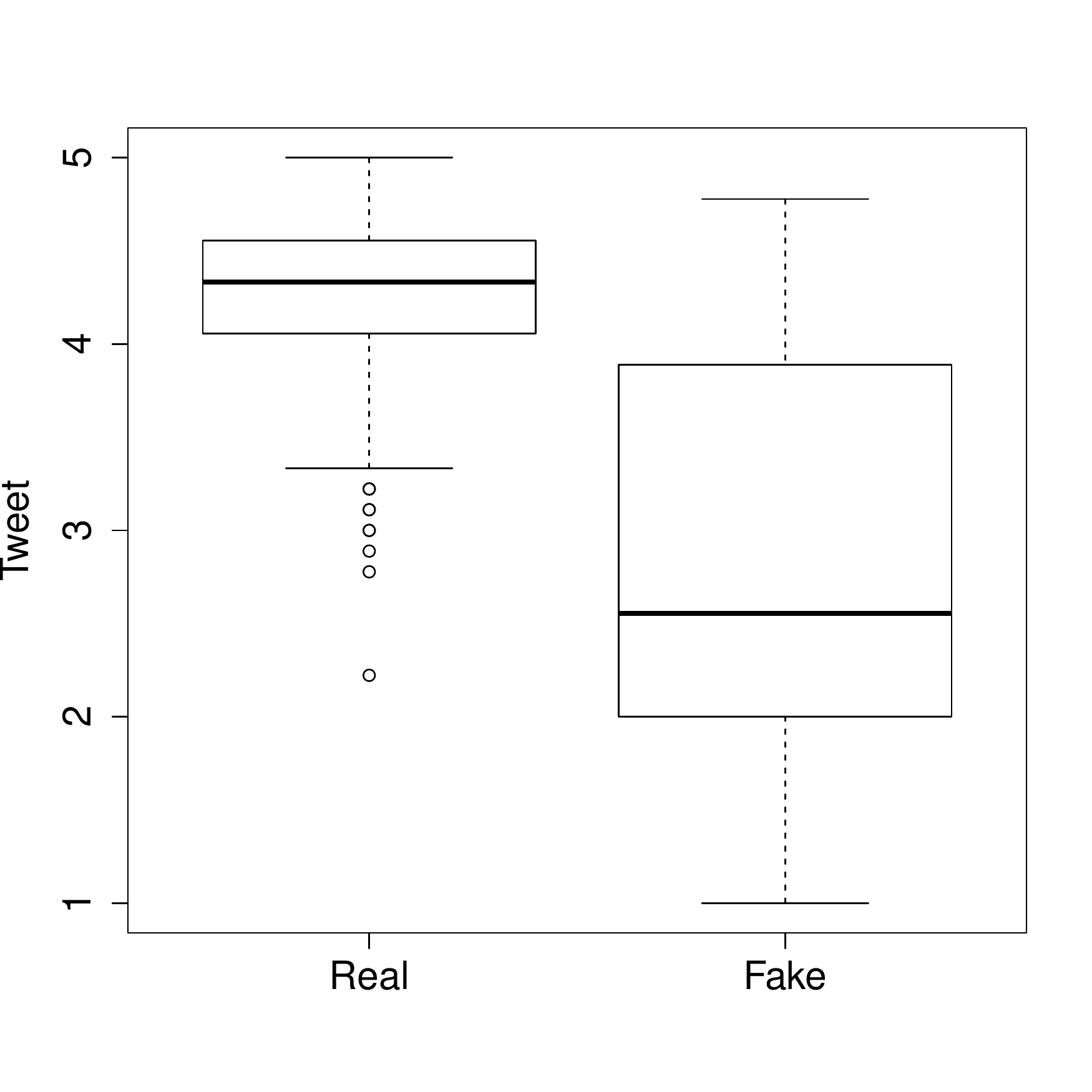}
  \caption{Tweet}
  \label{fig:f-box}
 \end{subfigure}
 \caption{Box plots for distributions of ratings across real and fake pictures on each of the features and the whole tweet.}
 \label{fig:boxplots}
\end{figure*}

Figure \ref{fig:boxplots} shows the distribution of ratings on different features across real and fake pictures. While all 6 graphs show that ratings were higher for real pictures, there are also fake pictures that were rated high in all cases, as well as real pictures rated low, showing the difficulty of the task. When looking at graphs separately, we see that ratings based on presentation (see Figure \ref{fig:e-box}) hardly help separate real from fake. After all, all tweets look the same in terms of design and layout, and just relying on how the tweet is written does not seem to be helpful to this end. Especially in the context of a natural disaster, where even real tweets might be quickly posted, probably from mobile devices, authors might not focus on presentation. It also stands out that corroboration shows very high ratings, both for real and fake pictures (Figure \ref{fig:b-box}) --we explain this later. When we look at authority ratings (Figure \ref{fig:a-box}), we notice that it received the lowest top ratings for fake pictures, which shows that users with authority did not hoax, and thus it seems to be a useful feature when it comes to verification of tweets. Both pictures (Figure \ref{fig:c-box}) and tweets (Figure \ref{fig:f-box}) received very high ratings for some fake pictures --potentially due to the problem of pictures that despite not being manipulated, they were taken from another context such as in a different location and/or time, and thus were not associated with Hurricane Sandy. As we showed in the examples of pictures in Section \ref{ssec:data-collection}, differentiating real and fake pictures is often challenging, since many of them (both real and fake) appear suspicious. The fact that some of the fake pictures and whole tweets were rated high demonstrates the need to rethink the way tweets are shown, at least when the goal is to verify their contents. The distributions of ratings for text plausibility (Figure \ref{fig:d-box}) are comparable to those for pictures and tweets. However, the median of plausibility ratings on fake pictures is much higher, showing that the text of many tweets pointing to fake pictures looks plausible.

Besides looking at distributions, we then looked at how accurate ratings were determining the tweets that contained real or fake pictures. For evaluation purposes, we define that a tweet would have been deemed real when it exceeds an average rating of 3.5, i.e., significantly higher than what we defined as uncertain (3). Table \ref{tab:accuracy-features} shows precision, recall, and F1 values. These results show that the way that most real tweets were correctly identified (in terms of recall) is by looking at the whole tweet. However, more than 16\% of tweets deemed real were actually fake. Interestingly, evaluations would have been slightly more accurate (in terms of precision) by looking only at the author. In fact, from the 39 tweets that workers mistakenly rated as real when looking at the whole tweet, 31 (79.5\%) were correctly identified as fake when just looking at the author. This matches up with findings by \cite{eagly1993psychology}, who revealed that the communicator is a strong feature to look at especially when the message itself is not easy to be evaluated. Nonetheless, looking just at the credibility of the author would miss many real tweets from authors who apparently did not look plausible. Looking at the authors of tweets, moreover, they were more accurate than just looking at pictures, while the latter could be expected to better describe a specific tweet when assessing its veracity. This confirms the conjecture by \cite{morris2012tweeting} that author information is currently underutilized when assessing credibility of tweets and could have been helpful for verification purposes. From the results, we can also confirm that presentation should not be considered to determine its credibility, as shown by its low accuracy, which would barely improve random decisions. A well-written and presented tweet is almost as likely to be fake.

It is also interesting to compare the accuracy of text plausibility and corroboration, as they are tightly related. The latter containing complementary information to that shown when rating plausibility, workers were slightly less accurate and did not achieve the improvement that could have been expected. By looking into more detail at corroboration ratings, workers surprisingly rated 71\% of fake tweets with a higher rating for corroboration than for text plausibility. Previously, psychologists stated that repeated exposure to a statement increases its acceptance as true, due to the fact that repetition of a plausible statement increases a person's belief in the referential validity or truth of that statement \cite{allport1945wartime,hasher1977frequency}. Our results buttress this fact by showing similar behavior on credibility perceptions from tweets. Regardless of the tweet being fake, reading the same claim repeatedly convinces users, misleading on many occasions.

\begin{table}[htb]
\begin{center}
 \begin{tabular}{|l|c|c|c|}
  \hline
  & \textbf{P} & \textbf{R} & \textbf{F1} \\
  \hline
  \textbf{Authority} & \textbf{0.849} & 0.546 & 0.665 \\
  \hline
  \textbf{Plausibility} & 0.748 & 0.880 & 0.809 \\
  \hline
  \textbf{Picture} & 0.825 & 0.829 & 0.827 \\
  \hline
  \textbf{Corroboration} & 0.739 & 0.903 & 0.813 \\
  \hline
  \textbf{Presentation} & 0.674 & 0.583 & 0.625 \\
  \hline
  \textbf{Tweet} & 0.838 & \textbf{0.931} & \textbf{0.882} \\
  \hline
  \textbf{Random} & 0.651 & 0.5 & 0.565 \\
  \hline
 \end{tabular}
\end{center}
\caption{Accuracy of raters assessing credibility of tweets from different features.}
\label{tab:accuracy-features}
\end{table}

\begin{figure*}[t]
 \centering
 \begin{subfigure}[b]{0.3\textwidth}
  \centering
  \includegraphics[width=\textwidth]{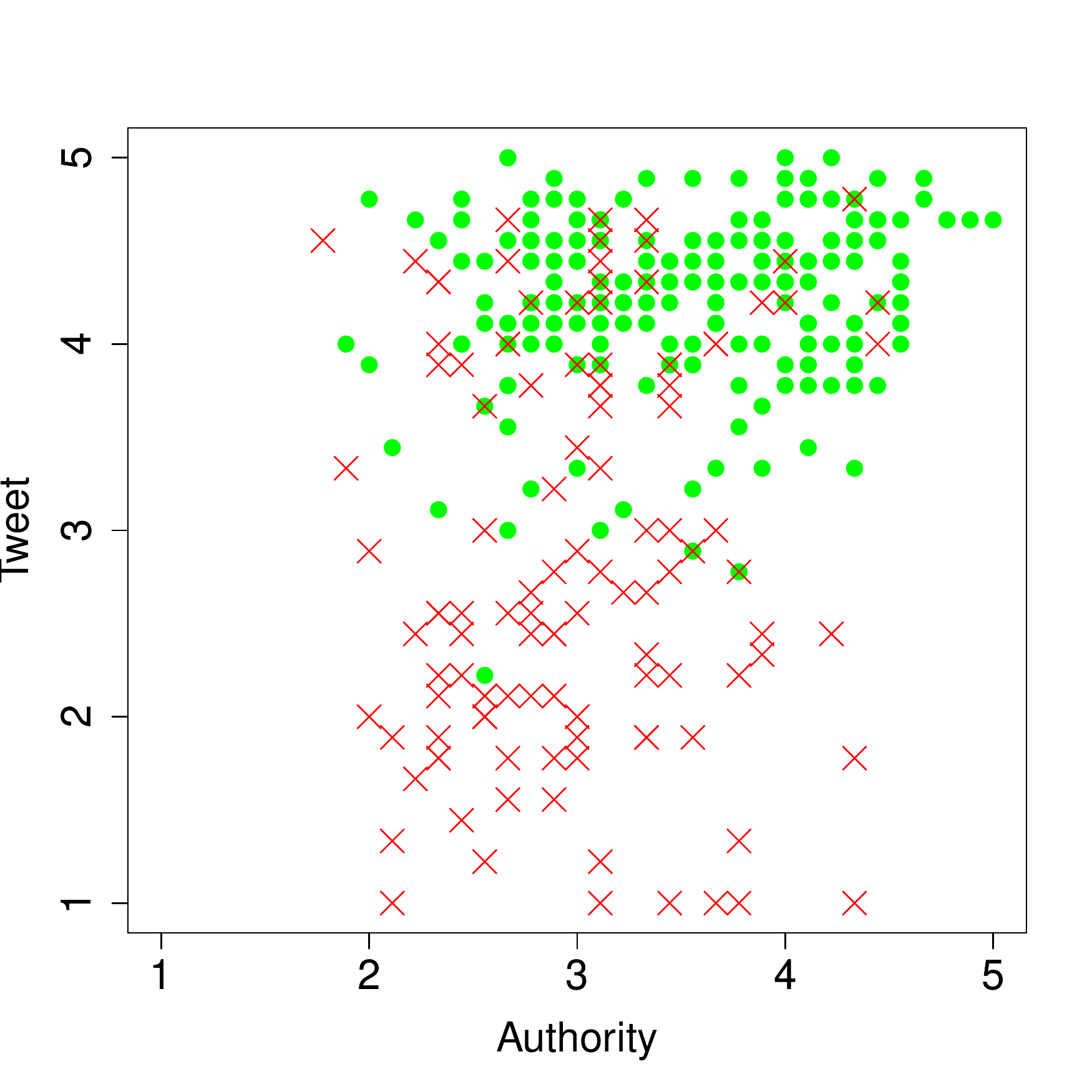}
  \caption{Authority (r = 0.34)}
  \label{fig:a-scat}
 \end{subfigure}
 \begin{subfigure}[b]{0.3\textwidth}
  \centering
  \includegraphics[width=\textwidth]{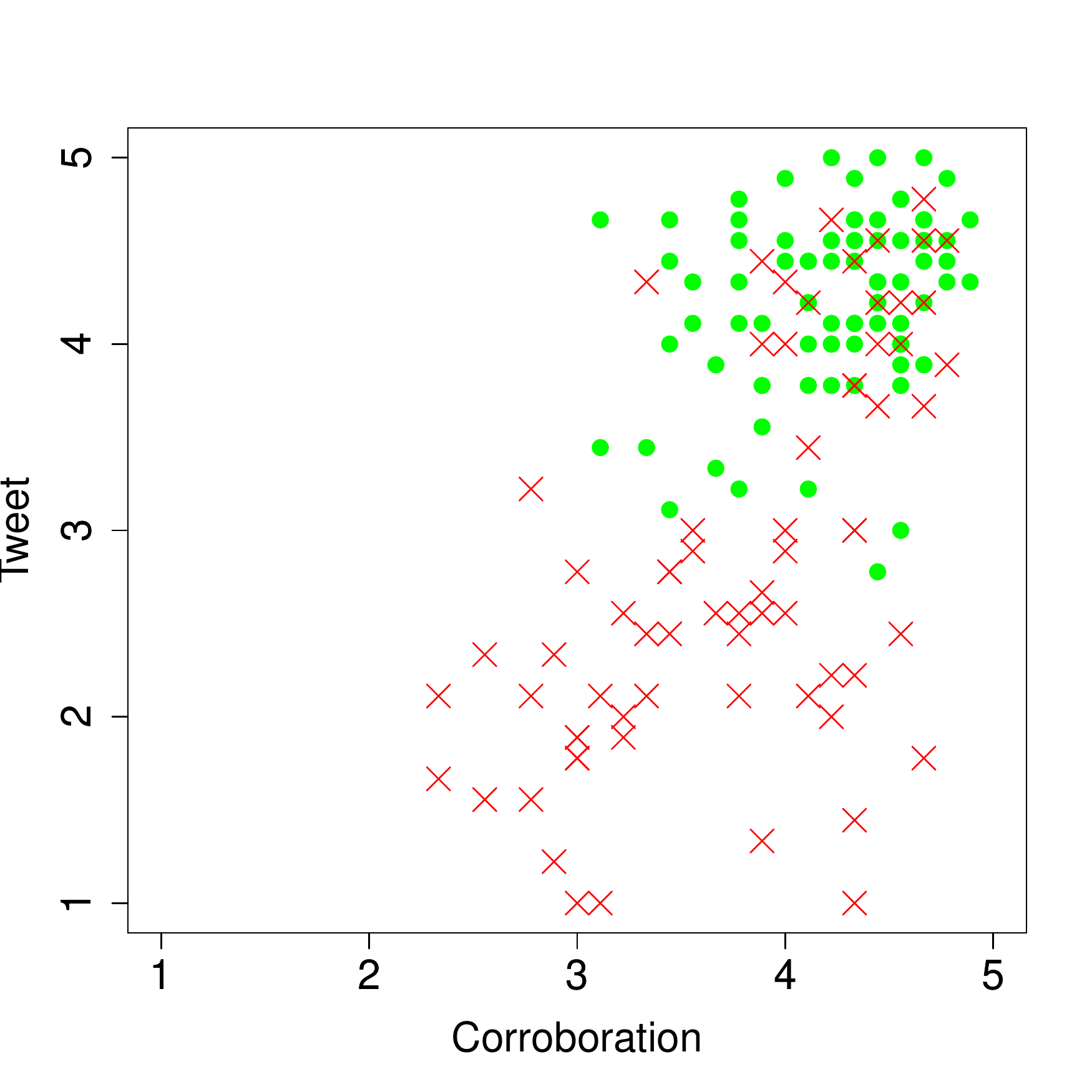}
  \caption{Corroboration (r = 0.61)}
  \label{fig:b-scat}
 \end{subfigure}
 \begin{subfigure}[b]{0.3\textwidth}
  \centering
  \includegraphics[width=\textwidth]{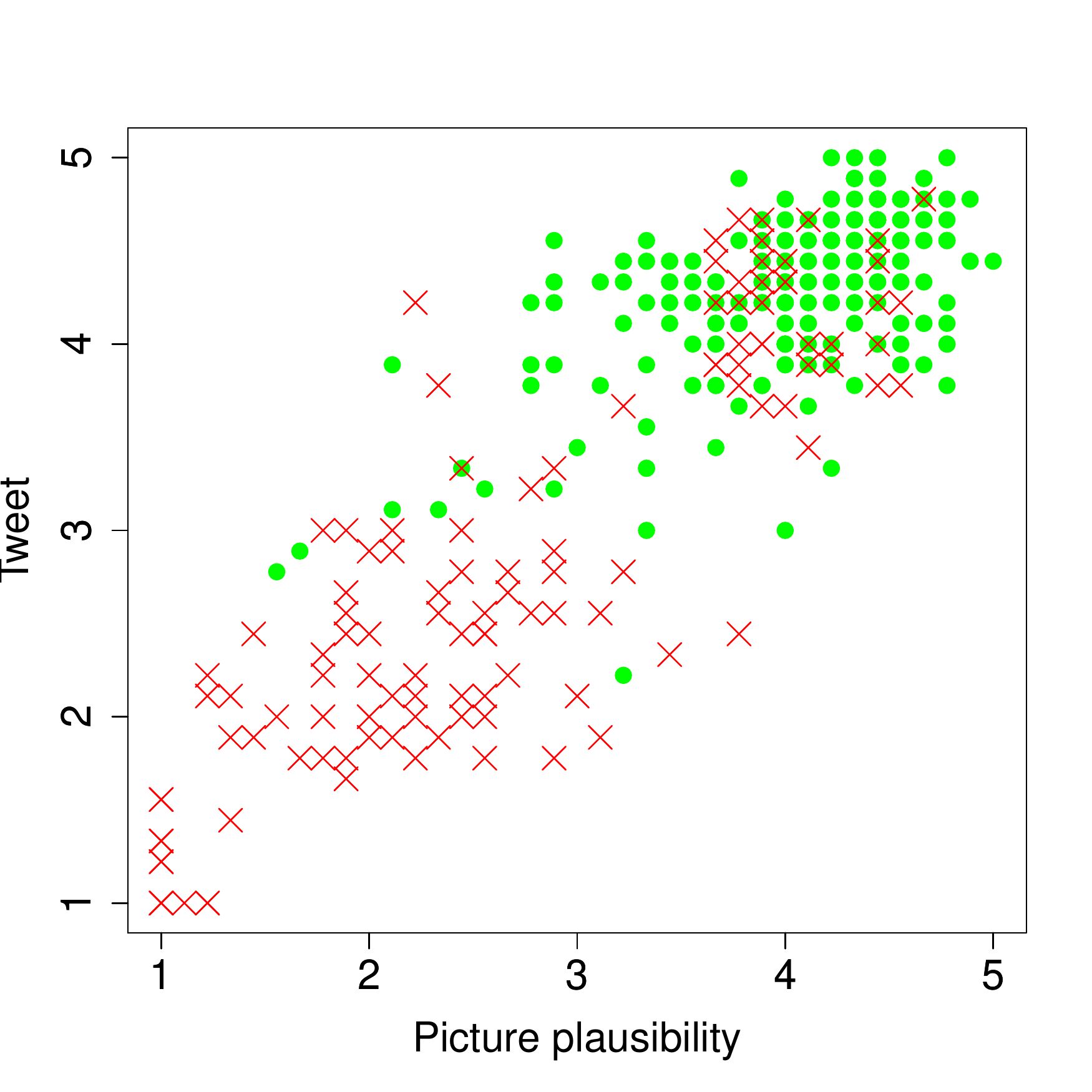}
  \caption{Picture plausibility (r = 0.87)}
  \label{fig:c-scat}
 \end{subfigure}
 \begin{subfigure}[b]{0.3\textwidth}
  \centering
  \includegraphics[width=\textwidth]{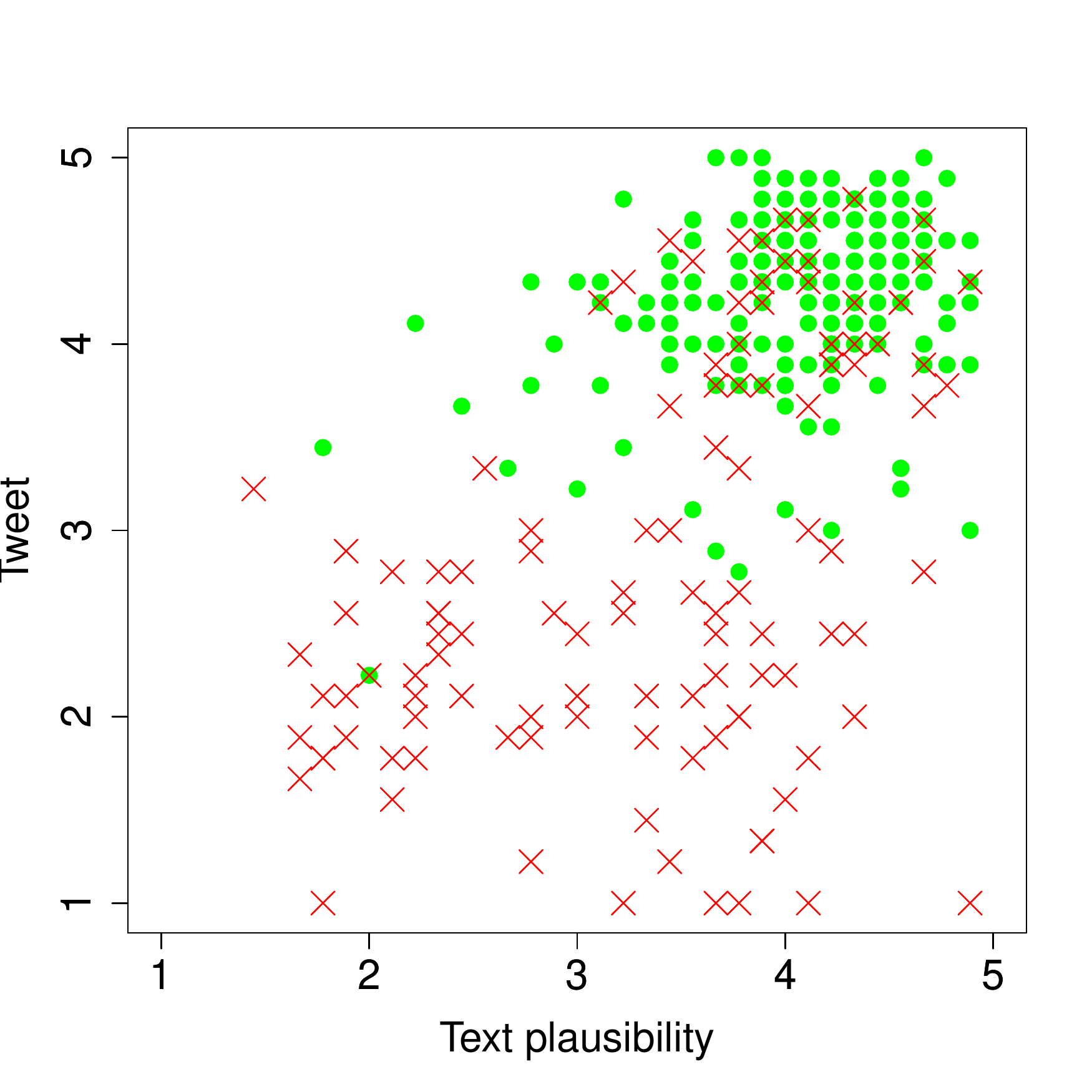}
  \caption{Text plausibility (r = 0.57)}
  \label{fig:d-scat}
 \end{subfigure}
 \begin{subfigure}[b]{0.3\textwidth}
  \centering
  \includegraphics[width=\textwidth]{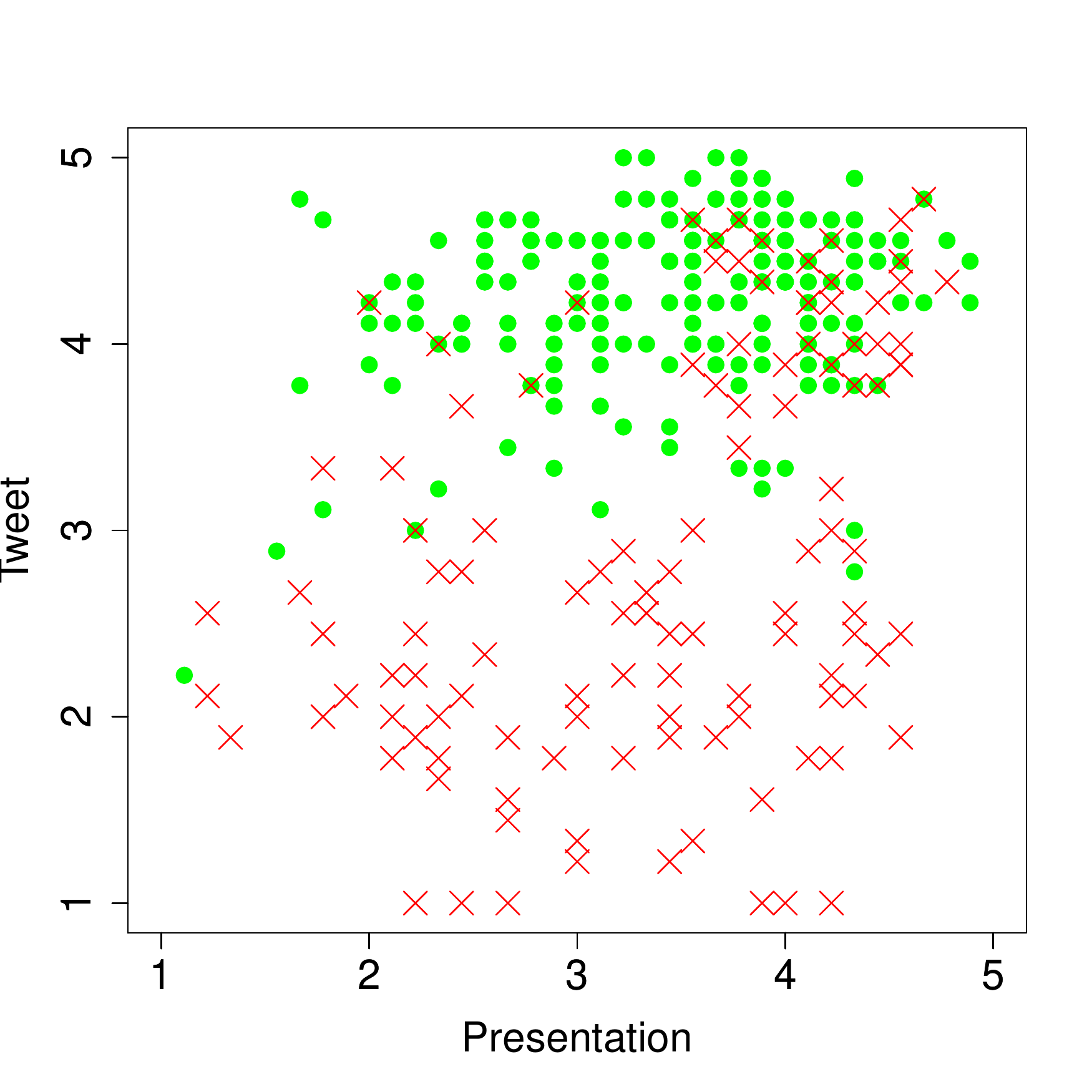}
  \caption{Presentation (r = 0.28)}
  \label{fig:e-scat}
 \end{subfigure}
 \caption{Scatter plots showing pairwise comparisons of ratings. All Pearson correlation coefficients (\textit{r}) are statistically significant ($p < .0001$). Red crosses refer to fake pictures, while green dots represent real pictures, as determined by professionals.}
 \label{fig:scatterplots}
\end{figure*}

We then looked at pairwise comparisons of ratings for each of the features to ratings of the whole tweet (see scatter plots in Figure \ref{fig:scatterplots}). This reinforces the fact that despite their high accuracy when analyzed separately, ratings on authority present very low correlation with ratings on whole tweets (Figure \ref{fig:a-scat}). Correlation with presentation ratings (Figure \ref{fig:e-scat}) is even lower, corroborating that it is not a reliable feature on Twitter. On the other hand, ratings on corroboration (Figure \ref{fig:b-scat}), picture plausibility (Figure \ref{fig:c-scat}) and text plausibility (Figure \ref{fig:d-scat}) show higher correlations with ratings on whole tweets, showing that users pay much more attention to the content of the tweet than its author --this is comparable to the finding by \cite{morris2012tweeting}, and suggests that more details from the author should be readily available to the user. Psychologists also suggest that the communicator's perceived credibility and expertise play an important in defining the persuasiveness of a message \cite{petty1986elaboration}, which is somewhat hidden at a first glance in a tweet. Having access to additional details of the author of a tweet at a first glance could have helped, especially in the context of a natural disaster, where checking the location and description of a user might provide further details.


\section{Discussion}

Our survey study analyzes credibility perceptions of tweets associated with pictures of Hurricane Sandy, and the accuracy of these perceptions with respect to the veracity of pictures as determined by professionals. The results obtained in this study suggest that visualizing more features along with each tweet would improve the verification process. Further, we shed light on improvements to be applied to a user interface focused on verification of tweets with the aim of increasing accuracy. While we are aware that features composing a tweet have to be analyzed altogether to make real assessments, our study collecting credibility perceptions on separate features and measuring their accuracy allows instead to understand how a careful analysis of each of the features helps toward assessing and verifying tweets. By asking AMT workers to rate features of a tweet separately, we guarantee that users do not get influenced by the rest of the components of a tweet.

We have found that having access to more details of the author of a tweet can be of help to assess the veracity of a tweet. In fact, the author is the feature that led the users to the most accurate perceptions (in terms of precision). Currently, only the Twitter handle and the profile picture are readily available on the site when reading a tweet, and showing other features such as the number of followers/followees, the location and the description of a user can help readers make better decisions. However, the author provides valuable details that need to be combined with other components of tweets for an efficient interpretation. We have also seen that decisions of users often relied on the picture associated to a tweet, which in some cases led to the wrong decision, and further looking at details of the author would have helped make more accurate decisions. Our results suggest that out of the 39 fake tweets mistakenly deemed real when workers looked only at the picture, 31 pictures (79.5\%) would have been correctly identified as fake when looking at author details. The importance of author details is therefore paramount on a social media service like Twitter, where counterfeiting plausible content is cheap, but building a reputable authority usually requires endeavor and longer time.

Moreover, we have seen that repeatedly seeing tweets supporting the same information, which we referred to as corroboration, can be harmful. Instead of clarifying and helping users make the right decision, seeing more tweets in the conversation often misleads the user to getting convinced, even when it comes to a fake claim. The presentation of a tweet, e.g., the spelling and grammar of a tweet, has proven useless toward verifying contents, different to what has been suggested for other information sources. The fact that many Twitter users might be using mobile devices when posting tweets increases the likelihood of making typos, disregarding capitalizations of letters, etc., which does not necessarily mean that the content of the tweet is fake.

Following the suggestions from epistemology researchers, an interface focused on verification for tweets could also benefit from incorporating other non-straightforward features. For instance, a score showing the extent to which a user can be considered reliable and has provided accurate information in the past would very likely be of help to assess the veracity of a tweet. A tweet from a reputable source that has always been accurate in the past could probably be considered truthful with no need to carefully analyze other features. This reputation score could be obtained either by collecting ratings from others on the site, or by automatically computing from its activity on the site. Currently, one way to infer the reputation of a user is by looking at the number of followers; however, this does not necessarily mean the user is reliable, as it might be for example a Twitter account created as a parody. The verification process would also benefit from strengthening some of the features that we have defined as not fully applicable to Twitter on this study. Some examples of features that could be strengthened include identification of the location of users, or being able to check if corroborations are independent to each other. Further research to cover missing aspects of the aforementioned features would be of help to enhance verification interfaces, albeit is not within the scope of this work.

On one hand, psychologists suggest that, since it is ultimately up to people themselves to decide whether to believe what they read on the Internet, people should certainly receive instruction about what features are indicative of accuracy \cite{fallis2004verifying}. In this regard, detailed guidelines on information verification from tweets would help users who consume the information. On the other hand, psychologists understand that people do not always apply the techniques for verifying the accuracy of information even when they know how. It can be expected that users will rarely investigate who the source of the information is even though they believe this to be an important indicator \cite{eysenbach2002consumers}. This is especially true for Twitter, where the stream of tweets flows rapidly, and little time is spent at reading each tweet. While showing additional user details along with tweets would help, users should receive instruction and be aware of the problem surfaced by the repeated exposure of fake information.

\section{Related Work}

Most of the research dealing with credibility on Twitter has focused on development of automatic techniques to assess credibility of tweets. \cite{castillo2011information} trained a supervised classifier to categorize tweets as credible or non-credible by using a set of predefined features they grouped in four types: message, user, topic, and propagation. They found the classifier to be highly accurate as compared to credibility assessments provided by AMT workers. Similarly, others have presented their research on automated classifiers or ranking systems by using graph-based methods \cite{gupta2012evaluating,wang2011bayesian,huang2012tweet}, using external sources such as Wikipedia \cite{suzuki2011credibility}, using content features \cite{odonovan2012credibility}, or comparing some of the previous methods \cite{kang2012modeling}.

When it comes to the study of credibility perceptions of users, the emergence of the Internet evoked an increase of interest on information credibility \cite{flanagin2000perceptions}. Researchers have suggested Web-specific features like visual design or incoming links \cite{hargittai2010trust,fogg2003users,lazar2007understanding}, and have given advice for assessing credibility of online content to avoid hoaxes \cite{piper2000better,koohang2003misinformation}. Researchers have also studied credibility issues in blogs and microblogs. \cite{johnson2004wag} surveyed blog users to rate credibility of blogs as compared to traditional media. They found that most users find blogs highly credible, and believe they provide more depth and more thoughtful analysis than traditional media. Regarding credibility of tweets, \cite{hermida2012tweets} examines the challenges presented by social media for tweet verification by journalists. He views Twitter as a media where information comes as a mix of news and information in real-time and with no established order, different from journalism's traditional individualistic and top-down ideology.

The first user study on credibility perceptions of tweets is that by \cite{morris2012tweeting}. Users were shown tweets with alterations from one another, such as different profile picture, or different tweet content. They studied how credibility ratings perceived by users varied according to these alterations. They found that the basic information shown on major tweet interfaces is not sufficient so as to assessing the credibility of a tweet, and that showing more details about the author of a tweet would help to that end. On a related study, \cite{yang2013microblog} conducted a survey study to compare credibility perceptions of U.S. users on Twitter and Chinese users on Weibo. They found cultural differences between both kinds of users, such as Chinese users being much more context-sensitive, and U.S. users perceiving microblogs generally as less credible.

Taking up a complementary goal to that by \cite{morris2012tweeting}, who explored the role of different features on credibility perceptions of tweets, our work studies how perceptions of tweets lead users to making a correct decision in a verification process aiming to identify truthful tweets and to get rid of hoaxes.

\section{Conclusion}

We have studied how credibility perceived from different features of tweets can help determine the veracity of a tweet in the context of a fast-paced event when the dearth of additional information sources makes it more difficult. We conducted for the first time an epistemic study of tweets, and studied how features suggested in epistemology can lead users to accurate or inaccurate information. We have surveyed users from the United States through Amazon Mechanical Turk on pictures posted on Twitter in the context of Hurricane Sandy. By analyzing the impact of components of tweets as to building accurate credibility perceptions, we have shed light on features that are not readily available on Twitter and most major applications, and could be of help to design a user interface that facilitates effective and efficient verification of contents from social media.

Our plans for future work include looking at complementary approaches to verification of digital content. These include for instance guidelines that journalists have put together to verify user generated content from social media \cite{silverman2014verification}. Other computational approaches that look at implicit features can also be of help to users evaluating social media content, e.g., looking at proximity of users and similarity of content in Twitter's social graph \cite{schaal2012analysis}.

\section*{Acknowledgments}

We would like to thank the anonymous reviewers for providing us with constructive comments and suggestions. This work was supported by the U.S. Army Research Laboratory under Cooperative Agreement No. W911NF-09-2-0053 (NS-CTA), U.S. NSF CAREER Award under Grant IIS-0953149, U.S. DARPA Award No. FA8750-13-2-0041 in the Deep Exploration and Filtering of Text" (DEFT) Program and Rensselaer Polytechnic Institute Start-up fund for Heng Ji. The views and conclusions contained in this document are those of the authors and should not be interpreted as representing the o cial policies, either expressed or implied, of the U.S. Government. The U.S. Government is authorized to reproduce and distribute reprints for Government purposes notwithstanding any copyright notation here on.

\bibliographystyle{spmpsci}      
\bibliography{si-deception-hoaxes}

\begin{thebibliography}{10}
\providecommand{\url}[1]{{#1}}
\providecommand{\urlprefix}{URL }
\expandafter\ifx\csname urlstyle\endcsname\relax
  \providecommand{\doi}[1]{DOI~\discretionary{}{}{}#1}\else
  \providecommand{\doi}{DOI~\discretionary{}{}{}\begingroup
  \urlstyle{rm}\Url}\fi

\bibitem{allport1945wartime}
Allport, F., Lepkin, M.: Wartime rumors of waste and special privilege: why
  some people believe them.
\newblock The Journal of Abnormal and Social Psychology \textbf{40}(1), 3
  (1945)

\bibitem{castillo2011information}
Castillo, C., Mendoza, M., Poblete, B.: Information credibility on twitter.
\newblock In: Proc. of the 20th international conference on World wide web, pp.
  675--684 (2011)

\bibitem{cotter2008influence}
Cotter, E.: Influence of emotional content and perceived relevance on spread of
  urban legends: A pilot study 1, 2.
\newblock Psychological reports \textbf{102}(2), 623--629 (2008)

\bibitem{eagly1993psychology}
Eagly, A., Chaiken, S.: The psychology of attitudes.
\newblock Harcourt Brace Jovanovich College Publishers (1993)

\bibitem{eysenbach2002consumers}
Eysenbach, G., K{\"o}hler, C.: How do consumers search for and appraise health
  information on the world wide web? qualitative study using focus groups,
  usability tests, and in-depth interviews.
\newblock Bmj \textbf{324}(7337), 573--577 (2002)

\bibitem{fallis2004verifying}
Fallis, D.: On verifying the accuracy of information: Philosophical
  perspectives.
\newblock Library trends \textbf{52}(3), 463--487 (2004)

\bibitem{fallis2008toward}
Fallis, D.: Toward an epistemology of wikipedia.
\newblock Journal of the American Society for Information Science and
  Technology \textbf{59}(10), 1662--1674 (2008)

\bibitem{flanagin2000perceptions}
Flanagin, A., Metzger, M.: Perceptions of internet information credibility.
\newblock Journalism \& Mass Communication Quarterly \textbf{77}(3), 515--540
  (2000)

\bibitem{fogg2003users}
Fogg, B., Soohoo, C., Danielson, D., Marable, L., Stanford, J., Tauber, E.: How
  do users evaluate the credibility of web sites?: a study with over 2,500
  participants.
\newblock In: Proc. of Designing for User Experiences, pp. 1--15 (2003)

\bibitem{goldman1999knowledge}
Goldman, A.: Knowledge in a social world.
\newblock Clarendon Press Oxford (1999)

\bibitem{goldman1986epistemology}
Goldman, A.I.: Epistemology and Congnition.
\newblock Harvard University Press (1986)

\bibitem{gupta2012evaluating}
Gupta, M., Zhao, P., Han, J.: Evaluating event credibility on twitter.
\newblock In: Proc. of SDM 2012 (2012)

\bibitem{hargittai2010trust}
Hargittai, E., Fullerton, L., Menchen-Trevino, E., Thomas, K.: Trust online:
  Young adults’ evaluation of web content.
\newblock International Journal of Communication \textbf{4}(1), 468--494 (2010)

\bibitem{hasher1977frequency}
Hasher, L., Goldstein, D., Toppino, T.: Frequency and the conference of
  referential validity.
\newblock Journal of verbal learning and verbal behavior \textbf{16}(1),
  107--112 (1977)

\bibitem{hermida2012tweets}
Hermida, A.: Tweets and truth: Journalism as a discipline of collaborative
  verification.
\newblock Journalism Practice  (2012)

\bibitem{hu2012breaking}
Hu, M., Liu, S., Wei, F., Wu, Y., Stasko, J., Ma, K.L.: Breaking news on
  twitter.
\newblock In: Proceedings of the SIGCHI Conference on Human Factors in
  Computing Systems, CHI '12, pp. 2751--2754. ACM, New York, NY, USA (2012)

\bibitem{huang2012tweet}
Huang, H., Zubiaga, A., Ji, H., Deng, H., Wang, D., Le, H.K., Abdelzaher, T.F.,
  Han, J., Leung, A., Hancock, J., et~al.: Tweet ranking based on heterogeneous
  networks.
\newblock In: COLING, pp. 1239--1256 (2012)

\bibitem{hume2001enquiry}
Hume, D.: An enquiry concerning human understanding, vol.~3.
\newblock Oxford University Press, USA (2001)

\bibitem{johnson2004wag}
Johnson, T., Kaye, B.: Wag the blog: How reliance on traditional media and the
  internet influence credibility perceptions of weblogs among blog users.
\newblock Journalism \& Mass Communication Quarterly \textbf{81}(3), 622--642
  (2004)

\bibitem{kang2012modeling}
Kang, B., O'Donovan, J., H{\"o}llerer, T.: Modeling topic specific credibility
  on twitter.
\newblock In: Proc. of the 2012 ACM international conference on Intelligent
  User Interfaces, pp. 179--188 (2012)

\bibitem{koohang2003misinformation}
Koohang, A., Weiss, E.: Misinformation: toward creating a prevention framework.
\newblock Information Science  (2003)

\bibitem{krippendorff2012content}
Krippendorff, K.: Content analysis: An introduction to its methodology.
\newblock Sage (2012)

\bibitem{kwak2010twitter}
Kwak, H., Lee, C., Park, H., Moon, S.: What is twitter, a social network or a
  news media?
\newblock In: Proc. of the 19th international conference on World wide web, pp.
  591--600 (2010)

\bibitem{landis1977measurement}
Landis, J.R., Koch, G.G.: The measurement of observer agreement for categorical
  data.
\newblock biometrics pp. 159--174 (1977)

\bibitem{lazar2007understanding}
Lazar, J., Meiselwitz, G., Feng, J.: Understanding web credibility: a synthesis
  of the research literature.
\newblock Now Publishers Inc (2007)

\bibitem{lewandowsky2012misinformation}
Lewandowsky, S., Ecker, U., Seifert, C., Schwarz, N., Cook, J.: Misinformation
  and its correction: Continued influence and successful debiasing.
\newblock Psychological Science in the Public Interest \textbf{13}(3), 106--131
  (2012)

\bibitem{lipton1998epistemology}
Lipton, P.: The epistemology of testimony.
\newblock Studies in History and Philosophy of Science-Part A \textbf{29}(1),
  1--32 (1998)

\bibitem{matheson2004weblogs}
Matheson, D.: Weblogs and the epistemology of the news: some trends in online
  journalism.
\newblock New media \& society \textbf{6}(4), 443--468 (2004)

\bibitem{morris2012tweeting}
Morris, M., Counts, S., Roseway, A., Hoff, A., Schwarz, J.: Tweeting is
  believing?: Understanding microblog credibility perceptions.
\newblock In: Proc. of CSCW 2012, pp. 441--450 (2012)

\bibitem{odonovan2012credibility}
ODonovan, J., Kang, B., Meyer, G., Hollerer, T., Adalii, S.: Credibility in
  context: An analysis of feature distributions in twitter.
\newblock In: SocialCom/PASSAT 2012, pp. 293--301 (2012)

\bibitem{o2012exposing}
O’Brien, J., Farid, H.: Exposing photo manipulation with inconsistent
  reflections.
\newblock ACM Transactions on Graphics \textbf{31}(1), 4 (2012)

\bibitem{petty1986elaboration}
Petty, R., Cacioppo, J.: The elaboration likelihood model of persuasion.
\newblock Advances in experimental social psychology \textbf{19}(1), 123--205
  (1986)

\bibitem{piaget1972psychology}
Piaget, J., Wells, P.: Psychology and epistemology: Towards a theory of
  knowledge.
\newblock Penguin Harmondsworth (1972)

\bibitem{piper2000better}
Piper, P.: Better read that again: Web hoaxes and misinformation.
\newblock Searcher \textbf{8}(8), 40--49 (2000)

\bibitem{schaal2012analysis}
Schaal, M., O’Donovan, J., Smyth, B.: An analysis of topical proximity in the
  twitter social graph.
\newblock In: K.~Aberer, A.~Flache, W.~Jager, L.~Liu, J.~Tang, C.~Guéret
  (eds.) Social Informatics, \emph{Lecture Notes in Computer Science}, vol.
  7710, pp. 232--245. Springer Berlin Heidelberg (2012)

\bibitem{silverman2014verification}
Silverman, C., Buttry, S., Wardle, C., Barot, T., Browne, M., Ingram, M.,
  Meier, P., Knight, S., Tsubaki, R.: Verification Handbook.
\newblock European Journalism Centre (2014)

\bibitem{stemler2001overview}
Stemler, S.: An overview of content analysis.
\newblock Practical assessment, research \& evaluation \textbf{7}(17), 137--146
  (2001)

\bibitem{suzuki2011credibility}
Suzuki, Y., Nadamoto, A.: Credibility assessment using wikipedia for messages
  on social network services.
\newblock In: Dependable, Autonomic and Secure Computing (DASC), 2011 IEEE
  Ninth International Conference on, pp. 887--894 (2011)

\bibitem{wang2011bayesian}
Wang, D., Abdelzaher, T., Ahmadi, H., Pasternack, J., Roth, D., Gupta, M., Han,
  J., Fatemieh, O., Le, H., Aggarwal, C.C.: On bayesian interpretation of
  fact-finding in information networks.
\newblock In: Information Fusion (FUSION), pp. 1--8 (2011)

\bibitem{yang2013microblog}
Yang, J., Counts, S., Morris, M., Hoff, A.: Microblog credibility perceptions:
  Comparing the united states and china.
\newblock In: Proc. of CHI 2013 (2013)

\bibitem{zubiaga2013curating}
Zubiaga, A., Ji, H., Knight, K.: Curating and contextualizing twitter stories
  to assist with social newsgathering.
\newblock In: Proc. of IUI 2013, International Conference on Intelligent User
  Interfaces (2013)

\end{thebibliography}

\end{document}